\documentstyle[12pt,aasms4]{article}

\begin{document}

\title{A RICH CLUSTER OF GALAXIES NEAR THE QUASAR B2 1335+28 AT 
z=1.1: COLOR DISTRIBUTION AND STAR-FORMATION PROPERTIES\footnote{Partly
based on observations made with the University of Hawaii 2.2 m telescope 
and the Isaac Newton Telescope. The INT is operated on the island of
La Palma by the Isaac Newton Group in the
Spanish Observatorio del Roque de los Muchachos of the Instituto de
Astrof\'\i sica de Canarias.}
} 

\author{Ichi Tanaka, Toru Yamada}
\affil{Astronomical Institute, Tohoku University, Aoba-ku, Sendai 980-8578, Japan; ichi,yamada@astr.tohoku.ac.jp}
\author{Alfonso Arag\'on-Salamanca, Tadayuki Kodama}
\affil{Institute of Astronomy, University of Cambridge, Madingley
Road, Cambridge CB3 0HA, U.K.; aas,kodama@ast.cam.ac.uk}
\author{Takamitsu Miyaji}
\affil{Astrophysikalisches Institut Potsdam, An Der Sternwarte 16,
Postdam, Germany}
\affil{MPE, Postf. 1603, D-85740, Garching, Germnay; miyaji@xray.mpe.mpg.de}
\author{Kouji Ohta}
\affil{Department of Astronomy, Faculty of Science, Kyoto University,
Sakyo-ku, Kyoto 606-8502, Japan; ohta@kusastro.kyoto-u.ac.jp}

\author{Nobuo Arimoto}
\affil{Institute of Astronomy, University of Tokyo, Mitaka, Tokyo 181-8566, Japan; arimoto@mtk.ioa.s.u-tokyo.ac.jp}

\begin{abstract}
We previously reported a significant clustering of red galaxies
($R-K=3.5$--$6$) around the radio-loud quasar B2 1335+28 at $z=1.086$
(Yamada et al. 1997, {\it Paper I\/}). In this paper, we establish the
existence of a rich cluster at the quasar redshift, and study the
properties of the cluster galaxies through further detailed analysis of
the photometric data.  We also list the positions, $K$-band magnitudes
and colors of all $K<19$ objects.  The near-infrared (NIR) $K$-band
imaging data presented in {\it Paper I\/}, together with some
additional $K$-band data, is newly analyzed  to study the extent of
the clustering of the red galaxies. We also constrain the cluster
redshift by applying a robust photometric redshift estimator and find a
strong peak around $z=1.1$.

The color distribution of the galaxies in the cluster is quite
broad and the fraction of blue galaxies ($\sim 70\%$) is much larger
than in intermediate-redshift clusters. With the help of evolutionary
synthesis models, we show that this color distribution can be explained
by galaxies with various amounts of star-formation activity mixed with
the old stellar populations.  Notably, there are about a dozen galaxies
which show very red optical-NIR colors but at the same time show
significant {\it UV\/} excess with respect to passive-evolution
models.  They can be interpreted as old early-type galaxies with a
small amount (a few $\%$ by mass) of star formation.  The fact that the
$UV$-excess red galaxies are more abundant than the quiescent red ones
suggests that a large fraction of old galaxies in this cluster are
still forming stars to some extent.  However, a sequence of quiescent
red galaxies is clearly identified on the $R-K$ versus $K$
color-magnitude (C-M) diagram.  The slope and zero point of their C-M
relation appear to be consistent with those expected for the precursors
of the C-M relation of present-day cluster ellipticals when observed at
$z=1.1$.  The scatter around the C-M line ($\sim 0.2\,$mag in $R-K$) is
twice as large as that of the morphologically-selected early-type
galaxies observed in rich clusters at $z < 1$, although the uncertainty
in the value of the scatter is quite large.  We estimate that the Abell
richness class of the cluster is $R_{\rm Abell}\sim 1$.  New X-ray data
presented here place an upper limit of $L_{\rm x} < 2\times
10^{44}\,{\rm erg\,s^{-1}}$ for the cluster luminosity. Finally, we
investigate the distribution of the galaxies over larger spatial scales
using our optical images, which cover a much larger area than the
near-infrared ones.  We find evidence that the cluster is located
within some lumpy over-dense structures, suggesting that the whole
system has not yet relaxed completely and is still dynamically young.

\end{abstract}

\section{Introduction}

Our knowledge of galaxy evolution in rich clusters at intermediate
redshifts ($z\lesssim 1$) has rapidly grown in the recent past thanks
both to important observational breakthroughs and the development and
improvement of galaxy formation and evolution models. It is now well
established that the majority of the early-type galaxies observed in nearby
and intermediate-redshift ($z\sim 0.5$) clusters contain  mainly old
stellar populations formed at high redshifts ($z_{\rm for}>2$; e.g.
O'Connell 1988; Bower et al.  1992; Arag\'on-Salamanca et al. 1993;
Rakos \& Schombert 1995; Gladders et al. 1998; Ellis et al. 1997;
Stanford et al. 1998; Kodama et al.  1998).

On the other hand, there is also extensive evidence suggesting that the
evolution of the entire cluster galaxy population is not as simple as
expected from monolithic formation scenarios. Such evidence is often
rather dramatic even at intermediate-redshifts:  the fraction of
luminous blue galaxies in rich cluster cores is known to increase
rapidly with redshift (Butcher \& Oemler 1978, 1984; Rakos \& Schombert
1995), and spectroscopically `active' (i.e., star-forming) galaxies
have been frequently observed in intermediate-redshift clusters (e.g.,
Dressler \& Gunn 1992; Couch \& Sharples 1987). Such blue and
star-forming galaxies are notably rare in $z\sim0$ rich clusters.
Recent ground-based spectroscopic and $HST$ high-resolution imaging
observations further revealed a population of old galaxies with
on-going or recent (within a few Gyrs prior to the observation)
star-formation activity as well as a significantly large fraction of
late-type disk galaxies and closely interacting systems (Abraham et
al.  1997; Morris et al. 1998; Oemler et al. 1997; Couch et al. 1998;
van Dokkum et al. 1998; Poggianti et al. 1999).  This `active'
evolution of the  cluster galaxies is possibly related to the evolution
of the morphological mix of the cluster galaxy populations (Dressler et
al. 1997; van Dokkum et al. 1998; Kuntschner \& Davis 1998).

An important step forward consists of studying clusters at higher
redshift, which can put stronger constraints on both the passive and
active evolution of the galaxies. A rapid change in the colors of
passively-evolving galaxies only occurs within $\sim2-3$ Gyr after the
end of the star formation (e.g., Bower et al. 1992). If cluster early-type
galaxies formed at $z>3$ and evolved passively, one can expect
conspicuous color changes only at $z\gtrsim 1$. At the same time, one
may expect more frequent star-formation activity at higher redshifts,
since there is a significant fraction of galaxies with {\it
post}-starburst signatures in intermediate-redshift clusters (Barger et
al. 1996; Postman et al. 1998; Poggianti et al. 1999).  A simple
extrapolation of the Butcher-Oemler effect (Butcher \& Oemler 1984)
also predicts a blue galaxy fraction  $> 50\%$ at $z\sim 1$. To date,
however, only a few rich clusters have been found at $z\gtrsim1$ and
studied in detail with multi-color photometry (Stanford et al. 1997;
Yamada et al. 1997), although the number of new candidates is rapidly
accumulating (e.g. Postman et al. 1996; Dickinson 1996; Olsen et al.
1998; Hall \& Green 1998; Ostrander et al. 1998).

Yamada et al. (1997, hereafter {\it Paper I}) revealed the
existence of a fairly rich cluster at $z\sim 1.1$ near the radio-loud
quasar (RLQ) B2 1335+28\footnote{In {\it Paper I}, we used the quasar
name Q~1335.8+2834 following H93. But this quasar was first listed in
the second B2 catalog of radio sources by Colla et al. (1972).
Following the recommendation by the IAU Commission 5 Task Group (chair,
H. R. Dickel), we use the original name for this quasar throughout the
paper.} through a deep NIR and optical imaging study.  The clustering
of galaxies in the field was first recognized by Hutchings et al.
(1993, hereafter H93). They found a $\sim 3\sigma$ number-density
excess of faint galaxies, many of which are blue in $R-I$, and claimed
that this region contains a very compact group of starburst galaxies.
They also detected 9 emission-line galaxies in the field using a
$75$\AA\ narrow-band filter centered on the wavelength of the
[OII]$\lambda$3727\AA\ line shifted to the quasar redshift.  In {\it
Paper I}, we showed that the field is characterized by the existence of
a significant number of very {\it red} objects in contrast with H93.
These red galaxies have optical-NIR colors consistent with those
expected for old galaxies at the quasar redshift. Existence of so many
red galaxies suggests that they are part of a fairly rich cluster at
high redshift, rather than just a group of star-forming galaxies as
suggested by H93.

In this paper we present the results of further analysis on this
cluster candidate.  We add new $K$-band data covering an area to the
East of the field previously analyzed in {\it Paper I}. These data
enable us to extend the region of the cluster for which a detailed
analysis based on $RIK$ photometric data is carried out.  At a first
glance, galaxies in the cluster region have a very wide color
distribution compared to those of nearby and intermediate-redshift
clusters, and no clear C-M relation is seen, except for a `red finger'
composed of the brightest red galaxies.  This may be due to
contamination by foreground objects and/or large photometric
uncertainties at fainter magnitudes.  However, it may also reflect an
intrinsic scatter in the stellar populations among the cluster member
galaxies such as the presence of different levels of on-going star
formation and/or age variations, since these would translate into a
significant color scatter.  We will show that the latter is the most
likely case in our cluster, i.e., the wide color distribution is
probably {\it intrinsic\/} and can be interpreted as various amounts of
star-formation activity in the cluster galaxies.

The paper is set out as follows. Section~2 describes the observations,
data reduction, object detection, number counts, and photometry.  A
catalog of $K$-selected objects is also given. Section~3 describes the
spatial extent of the candidate cluster galaxies, and the results of
photometric redshift estimates.  In section~4 we analyze the
optical-NIR colors of the $K$-selected sample.  The revised two-color
diagram and the C-M diagram are presented.  In section~5 we use the $R$
\& $I$ data, which cover wider areas, to investigate the properties of
the galaxy clustering on larger spatial scales.  Section~6 describes
the results of {\it ROSAT\/} HRI X-ray observations of the candidate
cluster.  In section~7, we estimate the cluster richness and blue
galaxy fraction. A brief comparison of our cluster with other
high-redshift ones is also given.  Section~8 summarizes the main
results of this paper.

\section{The Optical \& Near Infrared Data}

Most of the imaging data presented in this paper are the same as those
used in {\it Paper~I}, with the addition of a new $K$-band image
covering the eastern side of RLQ~B2~1335+28.  For convenience, we label
each region following Figure~\ref{hze_fig1}.  Since we have already
described the data acquisition and reduction procedures in {\it Paper~I}, 
we only include here a brief description, and add some improvements
carried out on the original reduction and analysis.  All the data were
reduced with the IRAF\footnote{The Image Reduction and Analysis
Facility (IRAF) is distributed by the National Optical Astronomy
Observatories, operated by the Association of Universities for Research
in Astronomy, Inc., under contact to the National Science Foundation.}
software packages.

\subsection{Optical Observations and Data Reduction}

The $R$- and $I$-band images were taken with the $1024\times1024$ TEK
CCD camera on the 2.5-m Issac Newton Telescope at the La Palma observatory
in February 1995. Total integration times were 8100s for the $R$ and 7500s for
the $I$-band images. The weather conditions were mostly photometric during
the observations, and the seeing conditions were variable, especially
during the $I$-band observations. The final co-added images covered an area
of $8.8 \times 8.8$ arcmin$^{2}$ ($900\times900$ pixel$^{2}$ with a 
pixel scale of 0.590 arcsec/pixel) and the stellar image sizes were 1.6
and 1.9 arcsec (FWHM) for the $R$- and $I$-band images, respectively.

In reducing the optical data, we followed standard CCD reduction
procedures. Flat fielding was done using sky-flat
frames. The photometric calibration was carried out using standard
stars from Landolt (1992) and the magnitudes were converted into
the Kron-Cousins photometric system. We neglected the Galactic
extinction given the high Galactic latitude of the field
($b=79.6^{\circ}$ and $N(H)=1.1 \times 10^{20}$ cm$^{-2}$; Burstein
\& Heiles 1982; Dickey \& Lockman 1990). Zero-point errors are estimated
to be smaller than 0.05 mag for the $R$-band data, but could be as large
as 0.1 mag for the $I$-band data.

\subsection{Near-Infrared Observations and Data Reduction}

The near-infrared (NIR) images were obtained with the University of
Hawaii 2.2-m telescope equipped with the QUIRC camera in February 1996.
A standard $K$-band filter was used. The detector was the $1024 \times
1024$ HgCdTe array with a $0.189$ arcsec/pixel scale. The F.O.V of the
camera is $3.22 \times 3.22$ arcmin$^{2}$. In order to make self-flat
frames, many dis-registered images with short exposures were taken in a
dither pattern of 8-arcsec separation.

We obtained $K$-band images for the two half-overlapped regions which
cover the western and eastern side of RLQ~B2~1335+28 (see Figure 2).
The western-field frame had a total exposure time of about 6000 sec and
was taken under fairly good seeing conditions. We reported the initial
results in {\it Paper I} using this frame (hereafter, {\it Paper I}
frame).  The eastern-field frame, however, was taken under poorer
conditions and it has two $\sim100\times100$ pixel$^{2}$ -sized ghosts
in the area. Only 3450 sec integration was achieved for this frame.  We
did not use these data (EAST frame) in {\it Paper I\/} because of its
relatively poorer quality, except when we investigate the spatial
distribution of the cluster.

The shapes of the stellar images in the EAST frame are somewhat
elongated. We compensated it by convolution with an ellipsoidal
Gaussian kernel. The resulting stellar image-sizes are $\sim2$
arcseconds (FWHM). One of the ghost patterns is located in the  region
which overlaps with the {\it Paper I} frame, and only the {\it Paper I}
frame is used in the overlapping area. There are only a few sources in
the ghost region and it does not affect any statistical results
discussed in the following sections. The other ghost image locates at
the lower-right (S-W) corner of the EAST frame and we excluded this
region in our analysis.

The {\it Paper I} and the EAST frames are reduced following a similar
procedure to that described in Gardner (1995) and are combined after
matching the stellar FWHM. As in Figure~\ref{hze_fig1}, we refer to their
overlapping area as the `Cluster' region, since the galaxy clustering
is most clearly seen there.  The East and the West sides of the Cluster
region are referred to as the `Eastern' region and the `Western'
region, respectively.

Magnitude calibration for the $K$-band data is done using three UKIRT
standard stars (taken from the on-line lists available at {\tt
http://www.jach.hawaii.edu/UKIRT/home.html}). The estimated zero-point
uncertainty is $\sim 0.04\,$mag.

Figure~\ref{hze_fig2} shows the $R$, $I$, $K$ false-color image of the
field. The quasar B2~1335+28 is the blue stellar object shown as
`QSO'.  The F.O.V. of this image corresponds to about $1.5\times2.3$
Mpc$^2$ at the quasar redshift\footnote{Throughout this paper, we
assume $H_{\circ}=50\,$km$\,$sec$^{-1}\,$Mpc$^{-1}$ and $q_{\circ}=0.5$.}.

\subsection{Object Detection and Photometry}

For the purpose of our optical-NIR color analysis, we first constructed
a catalog of $K$-band-selected objects (to avoid short-term
star-formation biases, cf. Arag\'on-Salamanca et al.  1993) and then
searched for their counterparts on the $R$- and $I$-band frames.
Object detection on the $K$-band frames was carried out using the 
SExtractor package (Bertin \& Arnouts 1996). The detection algorithm uses a
$\sigma=5$ Gaussian kernel and the parameters THRESHOLD$=0.6$ \&
MINAREA$=25$ pixel. The MAG\_BEST value in the SExtractor output is
used as a total magnitude for each object.  The Colors are measured in a
fixed-diameter aperture of 3.5 arcsec using the IRAF APPHOT package
(Davis 1987) after matching the FWHM of each image
(FWHM$\sim2\,$arcsec).

For the faint objects, the photometric errors arise mainly from the sky
fitting procedure.  We repeated the APPHOT photometry 7 times by
changing the radius and width of the sky annulus. By averaging these
measurements after excluding the minimum and the maximum values, we
obtained the final aperture magnitudes. Since the nominal APPHOT
magnitude errors are always smaller than the actual dispersion of the
repeated photometry, we used half the range of the magnitudes obtained
by this procedure as an indication of the individual photometric
error.

\subsection{Number Counts}
The differential number counts for each frame were used to estimate the
detection completeness.  Note that the source detection was carried out
independently on each frame.  Figures~\ref{hze_fig3}a and \ref{hze_fig3}b show
the results for the $R$- and $I$-band data, respectively.
Figures~\ref{hze_fig3}c and \ref{hze_fig3}d show the $K$-band counts for the
{\it Paper I} frame and the EAST frame, respectively (see caption).

From Figure~\ref{hze_fig3}a \& \ref{hze_fig3}b we estimated the nominal
detection completeness limit for the $R$- and the $I$-band frames to be
roughly 25$\,$mag and 23.5$\,$mag, respectively. For the $K$-band
frames, Figure~\ref{hze_fig3}c shows that the detection is complete to at
least the 19.5$\,$mag bin. The number counts for the Cluster region in
Figure~\ref{hze_fig3}c and \ref{hze_fig3}d agree very well down to $K = 19$.

Note that the counts for the Cluster region (filled circles) always exceed 
those of the Eastern region (open circles) in the magnitude ranges  
$K>17$, $R=21-23$, and $I=20-23$.

\subsection{The Catalog}

We present the catalog of the $K$-band-selected objects with $K<19$ in
Table~\ref{tbl-1}. Column (1) gives the object ID number.  The first
digit differentiates the region in which each object is located; 1, 2,
and 3 correspond to the Cluster region, the Western region, and the
Eastern region, respectively.  Column (2) \& (3) show the coordinates
of each object in arcsecs, relative to the quasar position, ie.
$(\alpha,\delta)=$ $(13^{\rm h}38^{\rm m}07^{\rm
s}.45,\;28^\circ\ 05^\prime\ 10^{\prime\prime}.7)$;
J2000 ---V\'eron-Cetty \& V\'eron 1996).  Positive offsets to the west
($x$) and to the north ($y$).  Column (4) gives the SExtractor MAG\_BEST
value.  The colors $R-I$, $R-K$, and $I-K$ and their uncertainties are
shown in Columns (5) -- (10).

\section{Clustering of Galaxies Near the Quasar B2 1335+28}

\subsection{Spatial Distribution of the $K$-Band-Selected Objects}
In Figure~\ref{hze_fig4}, we show the distribution of
all the objects with $K<19$ detected in the combined $K$-band frame.
Large filled circles, small filled circles, and open circles show
the objects with $R-K > 5$, $4 < R-K < 5$ and $R-K < 4$, respectively.

There is strong evidence for the existence of a fairly rich cluster
composed of more than 30 red galaxies brighter than $K=19$.  There are
only a few red galaxies beyond $\sim 20\,$arcsec East and $\sim
80\,$arcsec West from the quasar. The position of the quasar is offset
from the center of the red galaxy clustering.

Although the clustering is stronger for the redder objects, it is also
significant for relatively bluer objects ($R-K < 4$).  There are 21
galaxies with $R-K < 4$ and $17 < K < 19$ in the Cluster region while
only 8 and 11 in the Eastern and the Western region, respectively,
which means that the excess of the galaxies with {\it blue} opt-NIR
color is more than $3\sigma$.  The clustering feature seen in our
$K$-selected sample is consistent with that of H93 which reported a
clustering of emission-line galaxies and blue (in $R-I$) galaxies in
this region (see Figure 4 in H93).

The concentration of galaxies does not look very strong, however, even
for the reddest objects. An extended distribution of red galaxies near
the south-west edge of the field is also evident. These features may
indicate that the cluster is still dynamically young, but the limited
sky coverage and lack of radial velocity information prevent us from
making a conclusive statement. In section~5, we will come back to this
point and discuss it further using the optical data which covers a much
wider area on the sky.

\subsection{Cluster Redshift and Contamination by Foreground Objects}

As we reported in {\it Paper I}, the optical-NIR colors of the red
objects are consistent with those expected for old galaxies observed at
$z\sim1$. The spatial distribution of the emission-line objects in H93
is similar to that of red galaxies, which strongly supports the idea
that the detected emission-lines are indeed [OII]$\lambda$3727
redshifted to $z=1.1$ and the cluster is at the redshift of the quasar
B2~1335+28.

We have applied a photometric-redshift technique developed by Kodama,
Bell, \& Bower (1999) to the objects with $K < 19$ in the Cluster
region.  The code is specially designed to pick up cluster members at
high redshift based on an evolutionary population synthesis model.
High performance can be achieved when the photometric bands bracket
4000~\AA\ break, as it is the case here where $RIK$
pass-bands are just catching the break at $z=1.1$.

Figure~\ref{hze_fig5} shows the estimated photometric redshifts (solid
line) for the $K<19$ galaxies. As a reference, the observed redshift
distribution of field galaxies with $K<19$ studied by Cowie et al.
(1996) is also plotted  (dashed line). The strong peak near $z=1.1$ is
prominent. 50 galaxies out of total 105 galaxies are estimated to have
the most probable redshift between 0.9 and 1.3.  We estimate that the
uncertainty in the photometric redshifts could be as large as 0.2 due to
the lack of bluer passbands and the relatively large photometric errors
(see Kodama et al. 1999).  Considering these uncertainties, most of
these 50 galaxies are likely to have redshifts near $z=1.1$.  Strong
clustering of these objects on the sky, close to the QSO, further
supports this view.

From these redshift estimates and the narrow-band data of H93,
we conclude that a cluster of galaxy probably exists at $z=1.1$, although
further confirmation by spectroscopic observation is surely desirable.

In fact, the contamination by foreground objects is expected to be
small for the reddest objects ($R-K > 5$). Such red colors can be
expected only for old galaxies observed at $z\gtrsim1$.  Also, the
objects with intermediate colors ($4<R-K<5$) are not likely to be
foreground galaxies evolving passively at $z=0.4$--$0.8$, since they
are always too blue in $R-I$ color (see Section 4.2).

For the objects with the bluest colors ($R-K<4$), the
contamination may be larger.  For example, some galaxies at
$(x,y)\sim(60,-20)$ in Figure~\ref{hze_fig4} are too bright
($K\sim15$--$16$ mag) to be galaxies at $z=1.1$. It is possible,
however, that except for these bright galaxies, even
the galaxies in this color range can be intrinsically
blue galaxies at $z=1.1$.  In fact, many of the emission-line galaxies,
which are probably at $z=1.1$, have colors in this range and their
clustering properties look similar to that of the red galaxies.  In any
case, we will need spectroscopic confirmation of these results.

\section{Properties of the Cluster Galaxies}

\subsection{Color-Magnitude Diagram}
Figure~\ref{hze_fig6} shows $K$ versus $R-K$ C-M diagram. Filled
circles, open circles, and open boxes represent objects in the
Cluster region, the Western region, and the Eastern region,
respectively.  Compared to the C-M diagram presented in {\it Paper I},
the red sequence of the galaxies at $R-K \sim 5.5$ looks more
significant\footnote{We found a
$\sim 0.1$ mag calibration error in the $K$ magnitudes reported in {\it
Paper I} due to a software problem.}.

However, the distribution of optical-NIR color is relatively broad
compared with that of low- and intermediate-redshift clusters.  No
clear C-M relation can be seen except for a `red finger' seen at $K
\sim 17$--$18$ and $R-K \sim 5.5$. This trend is not so
atypical for clusters at high redshifts.  Although Stanford et al.
(1998) estimated the dispersion of the C-M relation to be as small as
0.1 mag for the {\it morphologically-selected\/} early-type galaxies in
clusters up to $z \sim 0.9$, the overall color distribution of the
entire sample of the cluster galaxies at $z\sim0.9$ is much broader (see also
Arag\'on-Salamanca et al. 1993).  A cluster of galaxies at $z=1.27$
discovered by Stanford et al. (1997) also has a very broad color
distribution except for the `red finger'.  The color distribution of
galaxies in the two clusters at $z\sim0.9$ studied in
Postman et al. (1998) and the 3C 324 cluster at $z \sim 1.2$ in
Dickinson (1996)  also have similar characteristics.

In Figure~\ref{hze_fig6}, we show the color tracks of passively-evolving
galaxies using the population synthesis model by Kodama \&
Arimoto (1997; hereafter KA97). The metallicity-sequence model for early-type
galaxies calibrated to the Coma cluster C-M relation (hereafter
``Coma C-M model'') is adopted for reference.
Thick solid curves show the evolutionary tracks of two model galaxies
with different luminosities ($M_{V}=-22$ and $-18.5$ mag
when they evolve into $z=0$) formed at redshift $z_{\rm f}=4.4$.
The tilted lines connecting the two galaxies mimic the predicted C-M
sequence at redshifts $z=0.2$, 0.6, 0.9, 1.1, and 1.4.
The tracks for different $z_{\rm f}$ (2.4 and 1.6) are also plotted.

The observed `red envelope',  a sequence of the reddest objects
in the C-M diagram, roughly follows the predicted C-M relation for the
oldest galaxies ($z_{\rm f}=4.4$) seen at $z=1.1$. Therefore at least some
of galaxies in the cluster could be old quiescent galaxies,
which are naturally regarded as the progenitor of the very old cluster
ellipticals seen in intermediate redshift and present day clusters.

\subsection{Two-Color Diagram}

In Figure~\ref{hze_fig7} we show the $R-I$ versus $I-K$ diagram for all the
objects brighter than $K=19$ mag.  Symbols are the same as in
Figure~\ref{hze_fig6} except for the ones enclosed by large diamonds, which
indicate the emission-line galaxies from H93. Two large empty diamonds
are the emission-line objects with $K>19$.

The two model tracks show the same Coma C-M model used in the C-M
diagrams.  The other two tracks labeled as ``tau = 1 \& 4 model'' 
correspond to continuous star-formation models with
exponentially-decaying time scales ($\tau$) of 1~Gyr and 4~Gyr,
respectively. We assume that these models mimic the colors of normal
disk galaxies; Lilly et al. (1998) showed that the colors of
disk-dominated galaxies at $z=0.65-0.87$ are broadly fitted by a
similar models with $\tau$=5 Gyr. The formation epoch of the models is
set to $z_{\rm f}=4.4$, and each track shows the color evolution from $z=0$
to 1.4 (from blue to red). Tick marks correspond to $\Delta z=0.1$
intervals. To show the color of galaxies dominated
by on-going star-formation at
$z=1.1$, we also consider a model galaxy having a constant star formation
rate and the age of 0.5 Gyr (the asterisk labeled as ``pure 
0.5Gyr burst''). Colors of dwarf (G0 V -- M5 V) and giant (G5 III -- M6 III) 
stars (Johnson 1966; Bessell 1990) are also plotted for reference.

A dozen or so galaxies ($\sim40\%$ of those with $I-K > 3.5$)
have colors consistent with those expected for the passively-evolving
galaxies at $z\sim$1.1. On the other hand, bluer galaxies with $I-K < 3.5$
could be interpreted as continuously star-forming galaxies
observed at various redshifts. However, as the excess in surface number
density of galaxies in the Cluster region is still higher than
$4\sigma$ in this color range, we believe that many of them
could be blue cluster galaxies at $z\sim1.1$.
Note that many of the emission-line galaxies in H93 also have
similarly blue colors.

There are also red ($I-K>3.5$) galaxies which are bluer in $R-I$ than
the passively-evolving galaxies ($R-I<1.3$).  Their $R-I$ and $I-K$
colors cannot be reproduced either by passively-evolving galaxies or by
mildly evolving, continuously star-forming galaxies at any redshift.
The $I-K$ colors are too red for continuously star-forming models with
$\tau>1\,$Gyr for any given $R-I$ color; their NIR light must be
dominated by old stars and some amount of $UV$ excess is needed to
explain their observed blue $R-I$ colors.  Hereafter we refer this
population ($I-K \gtrsim 3.5$ and $R-I\lesssim1.3$) as `$UV$-excess red
galaxies'.

On the whole, the galaxies in the Cluster region constitute a broad
sequence from the `Coma C-M model' to the `recent $0.5\,$Gyr burst
model'.  We try to explain this sequence by adding different amounts of
a star-formation component to the old passively-evolving galaxy model
at $z=1.1$ (hereafter we will call the former the {\it SF\/} component
and the latter the {\it OLD\/} component).  The rest frame near-$UV$
color ($\sim R-I$ color at $z\sim1.1$) of the model galaxy is
determined by the ratio between the $SF$ component (star-formation
rate) and the $OLD$ component.  As a template spectrum of the {\it
SF\/} component, we adopt a constant-SFR model with solar metallicity,
Salpeter IMF (mass cutoffs M$_{l}=0.1\,$M$_{\odot}$ and
M$_{u}=60\,$M$_{\odot}$), observed $0.5\,$Gyr after the onset of star
formation. These {\it OLD\/}+{\it SF\/} models can also mimic the
colors of exponentially-decaying models observed at $z=1.1$. For
convenience in characterizing the amount of the {\it SF\/} component
relative to the {\it OLD\/} one, we introduce the burst strength,
$f_{\rm burst}$, which is defined here as
\begin{equation}
 f_{\rm burst}= \frac{SFR\cdot t_{\rm burst}}{M_{\rm s}}
\end{equation} The $f_{\rm burst}$ value represents the mass fraction
of the burst population to the total stellar mass $M_{\rm s}$ (i.e.,
{\it SF\/} and {\it OLD\/} component).  Note that this value depends on
the duration of the burst, $t_{\rm burst}$, which is arbitrary chosen
to be $t_{\rm burst}=0.5\,$Gyr here.  However, the location of the model
tracks on the two-color diagram does not change significantly when
varying $t_{\rm burst}$.

The results are shown in Figure~\ref{hze_fig8}.  Many galaxies in the
Cluster region seem to follow the model track\footnote{The photometric
redshift measurements presented in Section 3 are just another
representation of this fact; see Kodama et al. (1999).} from a pure
passively-evolving galaxy to a pure star-burst one. Thus, the broad color
distribution of the galaxies in the Cluster region can be explained by
the variation of the fraction of the $SF$ component in galaxies at
observed at $z=1.1$.

Many of the emission-line galaxies detected by H93 lie around the track
of large burst strengths, supporting the hypothesis that the emission
line is indeed the redshifted [OII]$\lambda 3727$\AA\ from the
star-forming galaxies in a cluster at $z\sim1.1$ (Kennicutt 1992).  The
[OII] emission-line flux may affect the $I$-band flux of objects at
$z=1.1$. We show the effect of an emission line with equivalent width
(EW) of 100\AA\ by an arrow on the bottom right-hand corner of
Figure~\ref{hze_fig7}. The effect of reddening is also shown, using the
extinction curve of Savage and Mathis (1979).  An emission line with an
EW of 100\AA\ is probably too extreme, judging from Figure~2 in H93.
However, an extinction of $A_{V}=0.5$--$1\,$mag may well be present in
star-forming galaxies at $z>1$ (e.g., Glazebrook et al. 1998; Sawicki \& Yee
1998).

Two red [OII] emission-line galaxies locate just on the color track of
the passive-evolution model at $z\sim1$.  According to H93, both have
EW([OII])$\,\sim50$\AA.  Considering that there is no sign of star
formation in their colors, these emission-lines may originate from AGN
activity. In fact, Dressler et al. (1985) found that some galaxies have
strong [OII] emission and red colors in clusters at $z<0.5$, all of them
identified as AGN (Dressler, Thompsom \& Schectman 1985; Kennicutt
1992).

The $UV$-excess red galaxies lie at $f_{\rm burst} \lesssim 0.03$, and
the amount of {\it SF\/} component required to explain the excess 
$UV$ light  seems to be small. It is worth noting that the fraction of
such object is very large in this cluster.  Among galaxies with $19 <
K$ and $I-K > 3.5$, more than 60$\%$ show the $UV$ excess. Although
Smail et al. (1998) have shown that a similar kind of $UV$-excess
galaxies exist in clusters at $z\sim0.2$, the fraction is comparatively
small (at most a few \%).

There are several possible interpretations for the relatively large
abundance of these $UV$-excess red galaxies.  If the star-formation
events occur intermittently in each individual galaxy (triggered, perhaps, by
infall of small gas clouds and/or gas-rich galaxies, or regulated by
some internal process such as refueling by stellar mass-loss), those
events must be very frequent.  The time interval for such events must
be comparable with the duration of each star-formation episode.
Another possibility is that the majority of the galaxies in the cluster
are still at the last stages of their formation. Their colors may be
explained by exponential models with $\tau$ smaller than $1\,$Gyr.  The
star-formation time scale for these galaxies may be short but the
star-formation tail may be still significant.

The $UV$-excess red galaxies may be related to the galaxies with larger
$f_{\rm burst}$ in a time sequence. Since strongly star-forming galaxies are
much rarer than quiescent early-types in low- to intermediate-redshift
clusters, the galaxies with large $f_{\rm burst}$ have to disappear by the
present epoch.  While they gradually cease their star formation and turn
into red quiescent galaxies, they may pass through the $UV$-excess
red (small $f_{\rm burst}$) phase.

It is worth considering the possible relationship between this on-going
star-formation activity in cluster galaxies at $z\sim 1.1$ and the {\it
post-starburst\/} features seen in many galaxies in clusters at lower
redshift. Spectroscopic studies have revealed the existence of a
significant fraction of post-starburst galaxies in clusters at
$z=0.3-0.5$ (e.g., Dressler \& Gunn 1992; Couch et al. 1994, 1998;
Poggianti et al. 1999). These post-sturburst galaxies must have
experienced some star formation at higher redshifts.  The weak fossils
of the past star-formation activity seen as relatively-strong Balmer
absorption lines in the spectra of the intermediate-redshift cluster
galaxies can be seen only during $\sim$ a few Gyr after the
star-formation episode ceased. It is thus not surprising that we see a
large fraction of star-forming galaxies instead of post-starburst
galaxies in high-redshift clusters at $z>0.5$.  The observed large
fraction of star-forming galaxies in the cluster near B2 1335+28 may
imply that such star-formation events are already active at $z\sim
1.1$.  Postman et al. (1998) also found another example of a cluster at
$z=0.9$ with a large fraction of star-forming galaxies. 

\subsection{Identifying the Color-Magnitude Sequence of the Cluster at 
z=1.1}

Our revised C-M diagram (Figure~\ref{hze_fig6}) shows only a broad
distribution of red galaxies but not a tight C-M relation. As seen in the
last subsection, the main reason for this wide distribution of colors
is likely to be the on-going star-formation events with various
strengths. Here we attempt to exclude galaxies having star formation
to retrieve a C-M sequence made of quiescent galaxies.

From the model calculation shown in Figure~\ref{hze_fig8},
we see that the effect
of star formation is to make the $R-I$ color bluer while the age and
metallicity differences do not affect this color so much. As seen in
Figure~\ref{hze_fig8}, the passive evolution track at $z \sim 1.1$ predicts
$R-I$ colors redder than 1.3. By selecting galaxies with $R-I > 1.3$,
we can thus effectively reject star-forming objects while this
introduces little bias for objects of various ages or metallicities. In
other words, with this criteria we can exclude the objects which have
{\it UV\/} excess with respect to the expected spectrum of
passively-evolving galaxies.

Figure~\ref{hze_fig9} shows the results (\ref{hze_fig9}a for $R-K$
versus $K$ and \ref{hze_fig9}b for $I-K$ versus $K$). Objects with
$R-I>1.3$ are marked by symbols with a small dot (hereafter we refer to
them as the {\it quiescent members}).  The dashed line shows the C-M
relation at $z=1.1$ predicted by the Coma C-M model with $z_{\rm
f}=4.4$.

A fairly well defined C-M sequence appears in Figure~\ref{hze_fig9}. It
seems to follow the expected C-M relation at $z\sim1.1$ about $\sim 3$
magnitude range and it is compatibly with being the precursor of the
C-M relation seen in low-redshift clusters. We applied a least square
fitting algorithm with 2-sigma rejection clipping (three outlying
galaxies are rejected) for Figure~\ref{hze_fig9}a and obtained a
best-fitting slope of $-0.01\pm0.1$.  We also estimated the r.m.s.
scatter using the biweight scale estimator after 2-sigma rejection
clipping (Beers, Flynn, \& Gebhard 1990). The resulting scatter is found to be
$0.22^{+0.05}_{-0.03}$ for $R-K$, which is significantly larger than the
photometric uncertainties. This scatter is about a factor of two larger
than that found for the morphologically-selected early-type galaxies in
a cluster at $z=0.895$ observed by Stanford et al. (1998).

There could be several causes for this relatively large scatter under
the assumption that the sequence defined by the {\it quiescent
members\/} is really a precursor of the C-M relation at lower
redshifts. Some small amount of star formation can remain, but the
effect is likely to be small, since the galaxies with significant star
formation were excluded with our $R-I$ selection cutoff. Another
possibility is a chance projection of two poorer clusters with
$\Delta z\sim 0.1$.  This possibility cannot be excluded without
spectroscopic redshifts.  And a certain amount of spread in the ages
of the {\it quiescent members\/} remains an interesting possibility.  We
can evaluate the age spread needed to explain the scatter by comparing
the observed colors with the models.  Note that the following arguments
are sensitive to the true cluster redshift and/or to systematic
zero-point errors in our photometry.  Independent photometry and
spectroscopic redshift measurement are needed before obtaining more
reliable values.

In Figure~\ref{hze_fig9}, we show the colors of passive evolution models
with various ages at the observed redshift. Thick lines are for a
metal-rich, bright galaxy model (M$_{V}=-22$ or $K\sim17\,$mag) and
thin lines for a metal-poor, faint one (M$_{V}=-18.5$ or
$K\sim21\,$mag).  Comparing with the observed colors, we see that the
mean stellar age of the passively-evolving galaxies in the cluster
seems to spread over a~$\sim1\,$Gyr range.  This amount of age spread
does not conflict with the tight C-M relations observed in clusters at
$z\lesssim0.5$. As shown in Figure~\ref{hze_fig6}, even the models with an
age difference of 2 Gyr at $z=1.1$ rapidly evolve and become
indistinguishable at $z=0.6$. The major star-formation epoch in these
{\it quiescent members\/} is likely to be $z_{\rm f}=2$--$5$ and
possibly not before $z_{\rm f}\sim5$ with the adopted cosmology.

\section{Spatial Distribution of the Optically-Selected Sample}

In this section we analyze the spatial distribution of galaxies 
selected from the optical CCD ($R$ \& $I$) data. The CCD frames cover a 
wider area on the sky (five times larger) than the $K$-band frames. As the
clustering of red galaxies seems to extend beyond $K$-band frames
(Section 3.1), the inspection of the galaxy distribution on larger
scales is necessary to constrain the whole extent of the cluster. 

We extract cluster galaxy candidates in a somewhat {\it a posteriori\/}
manner, using the information obtained in the previous discussion. We
choose galaxies with $R-I$ colors and $I$ magnitudes similar to those
of the {\it red\/} candidate cluster members found in the $K$-selected
sample.  Although some contamination by foreground galaxies and the
drop out of some bluer cluster galaxies are inevitable, this procedure
is likely to maximize the contrast of the putative cluster galaxies
over the field galaxies.

In Figure~\ref{hze_fig8} we see that many galaxies in the Cluster region
have colors $R-I > 0.8$  while few galaxies in the Western and the
Eastern regions locate in this color range. Passively-evolving old
galaxies at $z=1.1$ have colors $R-I > 1.3$.  The brightest [OII]
emitter satisfying the red color criteria has $I=21.1$ mag. Thus, we
select the objects which have colors and magnitudes $21 < I< 23.5$ and
$R-I  > 0.8$ for the cluster galaxy candidates and especially those
with $R-I  > 1.3$ for candidates of passively-evolving galaxies in the
cluster.  The limit of $I=23.5$ is chosen for reliable color measurements.

We plot the sky distribution of all galaxies with $I<23.5$ in
Figure~\ref{hze_fig10}.  The red objects with $21<I<23.5$ are marked by the
filled circles.  We can see that the red galaxies (large filled circles
for $R-I>1.3$ and small filled circles for $0.8<R-I<1.3$) are clearly
clustered around the coordinates ($0$,$-50$). There seems to be other
significant excess features near ($100$,$-100$) and ($180$,$-100$).
Since a part of this `southern extension' of the cluster is also seen
in the distribution of the $K$-selected galaxies and some of them have
red optical-NIR colors consistent with galaxies at $z=1.1$, we
speculate that these features are related and possibly connected with
each other in real space. Thus, the whole red galaxy clustering
seems to extend beyond the $K$-band area in the South-West direction and it
may be a fairly elongated structure.  There is another concentration of
faint red galaxies around (0, 250), far North from the quasar B2~1335+28;
H93 also comments on a slight galaxy excess North of the quasar.

\section{The {\it ROSAT } HRI Observation and Results}

With the aim of investigating the possible extended X-ray emission
associated with the cluster, we have observed this field with the {\it
ROSAT\/} High Resolution Imager (HRI, David et al. 1996).  A $44.0\,$ks
exposure centered on $(\alpha,\delta)=(13^{\rm h}~38^{\rm m}~31^{\rm
s}.0,\;28^\circ\ 06^\prime\ 00^{\prime\prime})$ (J2000) was made with B2
1335+28 located in the central region of the FOV (off-axis
angle$\,=5\arcmin.3$).  In our analysis, we have only used the HRI raw
pulse height channels between 2~and~8, where the loss of the cosmic
signal is minimal ($\sim 7\%$) and at the same time the particle
background can be significantly reduced (David et al. 1996). We have
also excluded the time intervals where the remaining background level
is high. A $35.6\,$ks total exposure remained after screening.

A faint point-like X-ray source was clearly seen at the position of the
quasar B2~1335+28 itself on the screened image, but no evidence for
extension of this source was present. We have further searched for
extended X-ray emission which might be associated with the cluster. A
total of 271 HRI counts were present within the $1\arcmin$ ($\sim
0.5\,{\rm Mpc}$) radius region centered on the G1 galaxy, excluding the
$10\arcsec$ radius region around the quasar. The background count
estimated from the surrounding annular region ($1\arcmin$--$5\arcmin$),
scaled by area, was 274.  The $1\arcmin$ region centered on the center
of light of the red galaxies ($R-K>4$) contained 276 HRI counts, while
the scaled background count was 268. In either case, no significant
X-ray enhancement was found.

These numbers lead to a $3\sigma$ upper-limit to the HRI count-rate in
the 1~arcminute-radius region of $1.4\times 10^{-3}\,{\rm
cts\,s^{-1}}$.  This upper-limit count-rate corresponds to a
0.5--2$\,$keV luminosity of $2\times 10^{44}\,{\rm erg\,s^{-1}}$,
assuming Raymond \& Smith plasma spectra (distributed as a part of
XSPEC; Arnaud 1996) with plasma temperatures appropriate for a galaxy
cluster ($2 \la kT\,{\rm keV} \la 8$) and chemical abundance of 0.3 (solar),
placed at $z=1.1$, and absorbed by our galaxy ($N_{\rm H}=1.1\times
10^{20}\,{\rm cm^{-2}}$; Burstein \& Heiles 1982; Dicky \& Lockman 1990).
The implications of this upper-limit luminosity are discussed in the
next section.

\section{Discussion: Cluster Properties}

\subsection{Cluster Richness} 
We will estimate the cluster richness from both the $K$-selected data
and the optical CCD data.  The most popular method to determine the
cluster richness is that defined by Abell (1958). Abell cluster
richness ($R_{\rm Abell}$) is the number of cluster galaxies within an
area of $1.5\,$Mpc radius and in the magnitude range $m_3$--$m_3$+2
($m_i$ is the magnitude of the $i$-th brightest galaxy in a cluster).
However, these values are easily affected by the evolutionary
properties of the cluster galaxies and by the assumed cosmology when
the clusters are at high redshift.  The uncertainty in the field
correction also becomes rather large at $z>1$.  Therefore the richness
values obtained here should be regarded as an indicative.

First, we estimate the richness using the $K$-selected sample.  The third
brightest galaxy among the {\it red} ($R-K>5$) ones has $K \sim 17.2$
mag. There are 69 galaxies in the Cluster region in the magnitude range
between 17.2 and 19.2$\,$mag and 15 and 19 galaxies in the Eastern and
the Western region, respectively. The net galaxy excess of the Cluster
region thus contains $\sim50$~galaxies, although this may be an
underestimate since the Eastern and the Western regions are still in
the outskirts of the cluster. The `nominal' value of the excess is
likely to be $\gtrsim 50$ and thus this cluster can be classified as
Abell richness class larger than 0, and probably 1.

We also estimate the cluster richness using the optical CCD data since
the $K$-band area may not be large enough to allow the field galaxy
correction to be carried out accurately.  The optical CCD data covers a
wider area, allowing us to obtain more accurate field counts within the
same frame. Here we use another measure of the richness, $N_{0.5}$, a
statistic introduced by Hill \& Lilly (1991). This is the count of
galaxies within a $0.5\,$Mpc radius around the cluster and in the
magnitude range between $m_{1}$ and $m_{1}+3$ (they used an $R$-band
magnitude).  They applied the N$_{0.5}$ analysis to the fields around
radio galaxies at $z\sim0.5$. Since the $R$-band at $z\sim0.5$ roughly
correspond to the $I$-band at $z\sim1$, we can make a direct comparison
with their measurements. We estimate $m_{1}$ for the cluster galaxies
to be $I=21$ mag.  As galaxy detection becomes incomplete beyond
$I>23$, we corrected the number counts using the $N(m)$ slope of Tyson
(1988).  Field galaxy counts are taken from the outer area of our
data.  The resulting $N_{0.5}$ count is $19$--$25$ for the richest
clustering around coordinates ($0$,$-50$) in Figure~\ref{hze_fig10}.
Following the richness calibration by Hill \& Lilly (1991), this
corresponds to an Abell cluster richness class 1 ($N_{0.5}=15\pm5$ for
$R_{\rm Abell}=1$), in good agreement with the richness estimated from
the $K$-selected sample.

The estimated richness ($R_{\rm abell}\simeq1$) is consistent with the
upper limit of the X-ray Luminosity ($L_{\rm x}\la 2\times 10^{44}\,
{\rm erg\,s^{-1}}$). Only $\approx 10\%$ of the $R_{\rm Abell} = 1$ and
$\approx 20\%$ of the $R_{\rm Abell}= 2$ clusters have X-ray
luminosities above this value (Briel \& Henry 1993).  Note that the
sensitivity of our {\it ROSAT\/} observation is such that it would
detect the X-ray emission from the Coma cluster ($L_{\rm x}\approx 5
\times 10^{44}\,{\rm erg\,s^{-1}}$, e.g. Briel, Henry, \& B\"ohringer
1992) placed at $z=1.1$, but it would fail to detect the Virgo cluster at
this redshift by a factor of several. An X-ray Multi-Mirror Mission
(XMM) observation on this field is planned with an order of magnitude
better sensitivity. If the $z=1.1$ cluster has an X-ray luminosity
comparable to the present-day $R_{\rm Abell}= 1$--$2$ clusters, it
should be detectable with the XMM observations.

\subsection{Blue Galaxy Fraction}

Next we examine the blue galaxy fraction of the cluster using both the
$K$-selected and $I$-selected samples.  To study the evolution of the
blue-galaxy fraction, Butcher \& Oemler (1984) introduced $f_{\rm
b}$, which is the fraction of bright bluer [$\Delta(B-V)<-0.2$ from the
central color of the C-M relation] galaxies within the R$_{30}$ radius
which includes 30\% of the total number of cluster members. The $f_{\rm
b}$ values are calculated for absolute-magnitude-limited samples (e.g.,
$M_{V}<-20$).  This limit is affected by the luminosity evolution of
the cluster galaxies, which makes it difficult to compare the $f_{\rm
b}$ values for low- and high-redshift clusters directly. It is also
difficult to apply the original criteria directly to our data. Butcher
\& Oemler (1984) used $B-V$, longwards of the 4000\AA\ break, while both
our $R$ \& $I$ bands sample shorter wavelength. Thus, we cannot apply the
same criterion of `blueness' as used in Butcher \& Oemler (1984).
Nevertheless, it is still interesting to quantify the fraction of blue
galaxies in an evolutionary sense, i.e., the fraction of galaxies which
show signs of star-formation activity when compared with the quiescent
members.

First, let us evaluate the blue galaxy fraction using the optical CCD
data.  We count galaxies within a 0.5~Mpc radius around the center of
the clustering and considered the magnitude range of $21<I<24$.  We set
the criterion for {\it red\/} galaxies to be $1.3 < R-I < 1.6$ and for
{\it blue\/} galaxies $0.4<R-I<1.3$ on the basis of the model
calculation shown in Figure~\ref{hze_fig8}. Data for the field correction
are taken from the outer area of our optical images. Note that although
this criterion does not correspond exactly with that of Butcher \&
Oemler (1984), it does separate very well quiescent and star-forming
galaxies (cf. Figure~\ref{hze_fig8}).  The resulting blue galaxy fraction
of the richest clustering component is $\sim60\%$, significantly higher
than that found in clusters at intermediate redshift.  Note, however,
that we are selecting galaxies at shorter rest-frame wavelengths than
those used by Butcher \& Oemler, and therefore our sample is biased
towards bluer galaxies when compared with theirs.

In order to estimate $f_{\rm b}$ using the $K$-selected ($K<19$)
galaxies, and because of the relatively small area covered by the
$K$-band data, we simply evaluated the fraction of the {\it blue\/}
galaxies to the total count in the Cluster region relative to the
Eastern \& Western regions.  As before, we call {\it blue\/} galaxies
those with $R-I<1.3$.  We count 65 galaxies in the Cluster region, and
15 and 14 galaxies in the Eastern and the Western regions (see
Figure~\ref{hze_fig8}).  On the other hand, there are 54 {\it blue\/}
galaxies in the Cluster region and 12 and 13 {\it blue\/} galaxies in
the Eastern and the Western regions. Thus, the nominal blue fraction of
the $K < 19$ sample is more than 80$\%$.  

Although these estimates are rather crude, they do seem to indicate a
very high blue fraction, roughly matching the numbers found for
clusters at $z\sim0.9$ (Postman et al. 1998; Lubin et al. 1998; Rakos
\& Schombert 1995). If confirmed, they would indicate that the
Butcher-Oemler effect continues to increase its strength towards higher
redshifts.

\subsection{Comparison with Other High-Redshift Clusters}

Let us now compare in some detail the photometric properties of our
cluster near B2~1335+28 with those of other high-redshift clusters.
CIG~J0848+4453 at $z=1.27$  (Stanford et al. 1997) is another good
example of $z > 1$ cluster whose photometric properties have been
published. It also shows a `red finger' similar to that of our cluster,
but the color distribution in $R-K$ of the `finger' galaxies seems to
be somewhat tighter. Moreover, their reddest envelope galaxies are
$\sim0.5\,$mag redder in $R-K$ than the predictions of the models
discussed in Section 4 when shifted to $z=1.27$. Considering that the
colors of the reddest galaxies in the cluster near B2~1335+28 are
marginally bluer than the model predictions, it seems as if the oldest
galaxies in CIG~J0848+4453 may be older than those in our cluster
(taking the redshift difference into account). Note, however, that
relatively small uncertainties in the photometric systems and
zero-points could accentuate the apparent differences, and it is
important to stress that the similarities are probably more significant
than the differences between these two clusters.

The optical-NIR colors of the `quiescent members' of our cluster show
some scatter, which we interpreted as possible {\it age differences\/}
among these passively-evolving galaxies.  Another example of a
large-scatter C-M relation in a high-$z$ cluster was very recently
found by Ben\'\i tez et al. (1998), who studied the cluster
AX~J2019+112 at $z=1.01$ using deep optical-NIR multi-color
photometry.  There are several galaxies which may define a C-M sequence
and they show significant  scatter in the $VIK$ two-color diagram,
preferentially on the $I-K$ axis. Further examples may be found in
da Costa et al. (1999).

Extended X-ray emission has been detected for the clusters
CIG~J0848+4453 (Stanford et al. 1997) and AX~J2019+112 (Ben\'\i tez et
al. 1998; Hattori et al. 1997). If the emission originates from the cluster hot
gas, it would suggest that these clusters are dynamically-evolved
systems.  Strong X-ray emission has not been detected from our cluster,
although the sensitivity limit of our data is rather marginal for a
cluster with Abell richness class 1, as discussed in \S6. The weak
concentration and the lumpy and elongated appearance of our cluster may
imply that it is dynamically too young to show strong X-ray emission.

The star-formation history of cluster galaxies may well be coupled to
the dynamical evolution of the cluster itself.  In a morphological
study of two $z\sim0.9$ clusters by Lubin et al. (1998) they found that
the fraction of early-type galaxies is higher in the more relaxed and
evolved system. In the companion paper, Postman et al. (1998)
calculated `color ages' for the galaxies from their $BVRI$ photometric
results.  The number of galaxies older than 4~Gyr (see Postman et al.
for details) is five for CL~1604, the rich and evolved cluster, while
none are found for the merging cluster CL~0023. Thus, it is not
surprising that in our cluster, which shows many signs of being
dynamically young, we also find a large fraction of galaxies with young
stellar populations and star-formation activity.

Finally, it is worth mentioning that the our cluster is located in a
somewhat unique environment. The quasar B2~1335+28 is one of the
members of the quasar concentration found at at $z=1.1$ by Crampton et
al. (1989).  Hutchings et al. (1995) found galaxy surface density
excess around other quasars in this concentration. If, as it seems
likely, the quasars trace an underlying supercluster of galaxies, our
cluster would be just a member of this (proto?-)supercluster.  It would
be extremely interesting to study the overall distribution of the
galaxies and their properties over much larger scales, covering the
whole quasar concentration.

\section{Summary}
In this paper we have presented an optical-NIR photometric analysis of
the field around the quasar B2~1335+28. Our main findings are: 

\begin{enumerate}

\item Using some newly-analyzed $K$-band imaging data, we have examined
the galaxy clustering over a larger area than that covered by
Yamada et al. (1997). We have confirmed the existence of a fairly rich
cluster of galaxies and determined that it extends  $\sim 90\,$arcsec
in the East-West direction.

\item The cluster contains galaxies with a wide range of colors, but we
argue that it is not likely to be heavily contaminated by a foreground
cluster. The large scatter of colors is probably intrinsic.

\item  Using photometric redshifts, we find a very strong peak at $z
\sim 1.1$. No spectroscopic redshifts are yet available. From the
spatial coincidence of the narrow-band-selected ([OII]-emitting)
galaxies and the $K$-selected candidate cluster galaxies, we conclude
that the redshift of the cluster is probably $z=1.1$, similar to that
of the quasar.

\item There is only a small fraction of galaxies which have colors
consistent with passive-evolution models. Substantial numbers of red
galaxies with some {\it UV\/} excess (bluer $R-I$ color) as well as bluer
galaxies with colors similar to those of mildly-evolving disk galaxy
models have been found. These bluer galaxies may be the precursors of
the galaxies with post-starburst features frequently seen in clusters
at intermediate redshifts.

\item The Abell richness of the cluster is $R_{\rm Abell} \sim 1$.
Although our {\it ROSAT\/} HRI data set an upper limit ($2\times
10^{44}\,{\rm erg\,s^{-1}}$) to any possible extended X-ray emission
from the cluster, this is still consistent with the estimated
richness.

\item The blue galaxy fraction is 60--80\%, substantially larger than
that of $z\lesssim0.5$ clusters, but comparable to other $z\sim1$
clusters. This suggests that the strength of the Butcher-Oemler effect
continues to grow with redshift.

\item  By selecting passively-evolving galaxies using a rest-frame {\it UV\/}
color criterion, we recover a clear optical-NIR color-magnitude
sequence which may be the precursor of the tight C-M relations seen in
lower-redshift clusters.  The C-M slope is consistent with
passive-evolution models, while its scatter is fairly large ($\sim
0.2\,$mag in $R-K$). The mean age of these galaxies is estimated to be
$\sim 2$ Gyr at the epoch of observation, although the absolute value
is somewhat uncertain due to the size of the photometric zero-point
errors.

\item Using $R$ and $I$-band images covering a wider field, we examine
the large-scale distribution of the galaxies with similarly red $R-I$
colors to those of the cluster galaxies in the $K$-selected sample.
Some `lumpy' and elongated structures are revealed, including an
extension of the cluster in the South-West direction. This result
suggests that the cluster is still dynamically young and may evolve
into a larger cluster in the future.

\end{enumerate}

\acknowledgments
This research was partially supported by grants-in-aid for scientific
research of the Japanese Ministry of Education, Science, Sports and
Culture (08740181, 09740168). Part of this work was also supported by
the Foundation for the Promotion of Astronomy of Japan.  We thank
E.F.~Bell and R.G.~Bower for allowing us to use the Kodama et al's (1999)
photometric redshift code in this paper.  
A.A.S.~acknowledges generous financial
support from the Royal Society.  T.K.~thanks the JSPS postdoctoral
fellowship for research abroad for financial support. T.M. thanks the
JSPS Japan-Germany Collaboration for High-Energy Astrophysics Program
for travel support to Tohoku University, where most of the X-ray
analysis was carried out. T.M.~also thanks Tohoku University for 
their hospitality during his visit.  K.O.~thanks the Institute of Astronomy,
University of Hawaii, where part of this work was done, for their
hospitality during his stay.  T.Y., A.A.S., and T.K. would like to thank the
International Program for Advanced Studies in Astrophysics ``Guillermo
Haro'', hosted by the Instituto Nacional de Astrof\'\i sica, \'Optica y
Electr\'onica (INAOE, Mexico) for their hospitality during the summer
of 1998, when many of the results presented here were discussed.


\newpage
\protect 
\figcaption[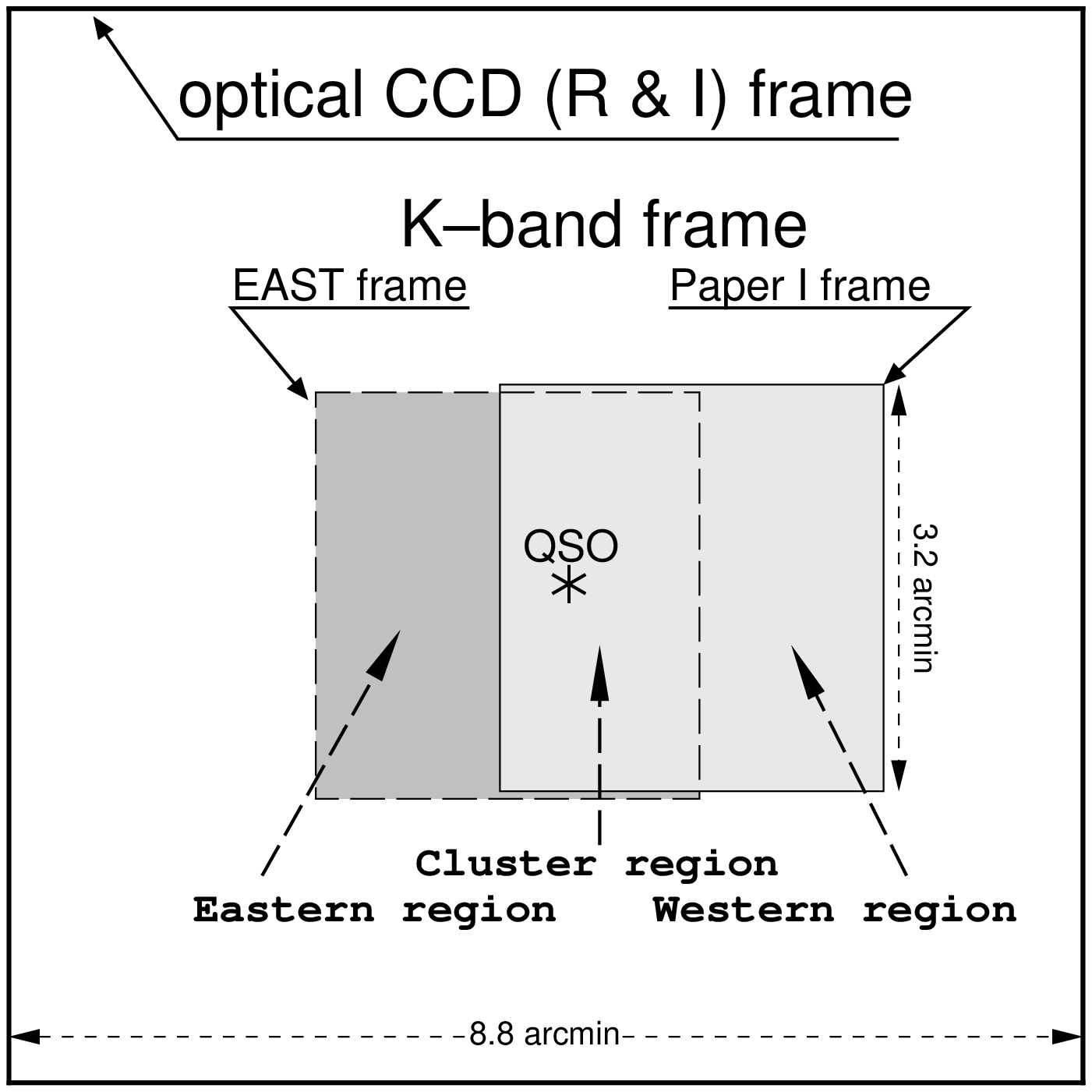]{A schematic illustration of the imaged fields. \label{hze_fig1}}

\figcaption{Three-color representation of the field around the quasar
B2~1335+28. The Blue, Green and Red images correspond
to the $R$-, $I$- and $K$-band
frames, respectively. North is up, and East is left. The absolute
magnitude of the quasar is $M_{\rm V}\sim-24$. The galaxy G1 is the reddest
and brightest galaxy in the cluster, and G2 is the reddest [OII]
emitter detected by Hutchings et al. (1993). A red patch in lower-left
corner is a ghost. \label{hze_fig2}}

\figcaption[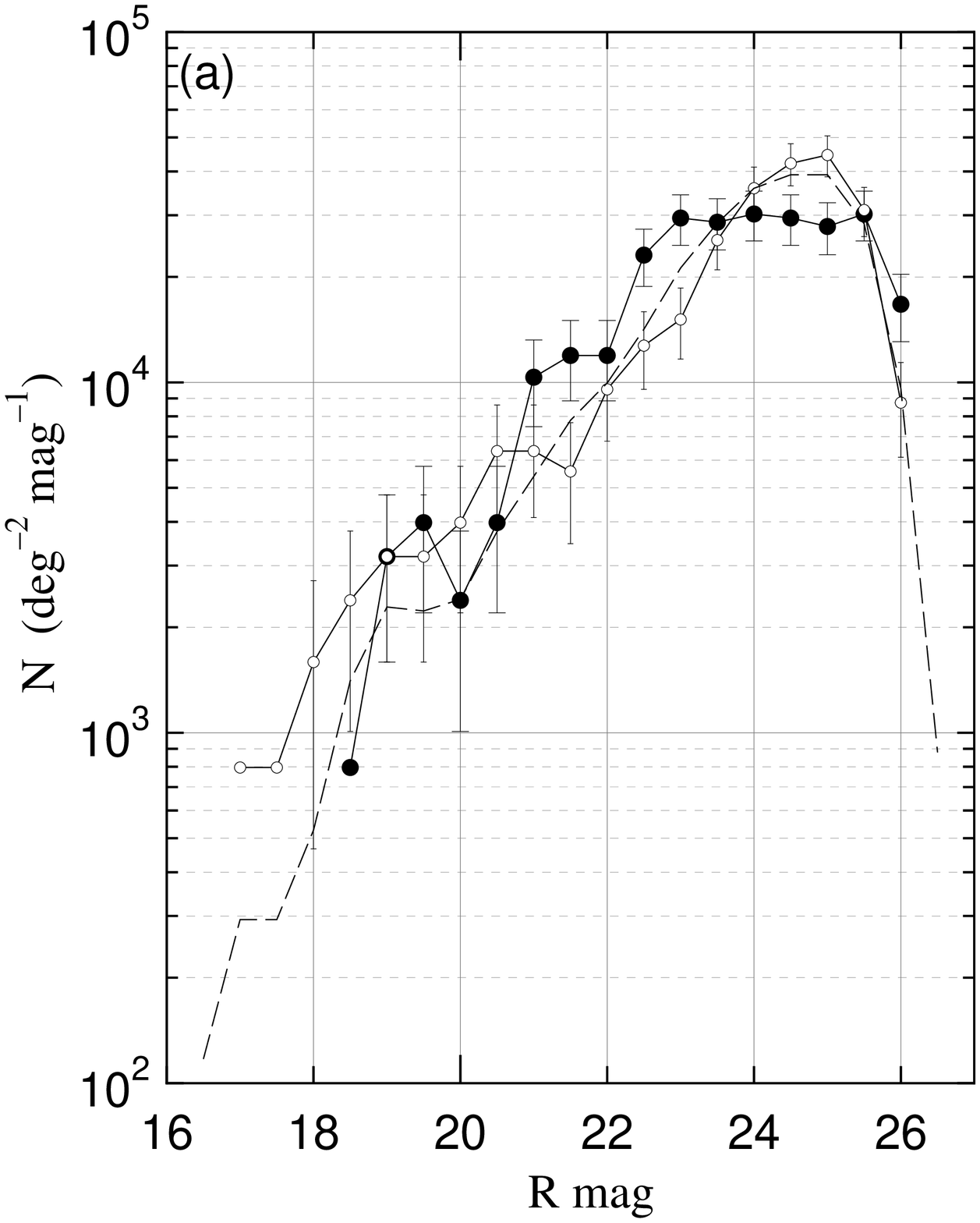]{The number counts of the objects
in each frame. Panel (a): The
$R$-band number counts. Panel (b): the $I$-band counts.  Filled and
open circles represent the counts in the Cluster region and the Western
region defined in the $K$-band frame (see Figure~\ref{hze_fig1}).  Dashed
lines in panel (a) \& (b) represent the counts in the complete optical
images excluding the area of the $K$-band {\it Paper I\/} frames and
around two bright stars. The ratios of these three areas are 1:1:13.
Stars are not rejected from these counts. Panel (c): $K$-band number
counts for the {\it Paper I\/} frame. Panel (d): $K$-band counts for
the EAST frame. Filled circles show the counts in the Cluster region
and open ones in the Western (c) and the Eastern (d) region. The dashed
line shows the $K$-band galaxy counts for the South Galactic Pole by
Minezaki et al. (1998). Stars are rejected from the counts in (c) and
(d). Comparing panels (c) and (d), we see that they are well matched to
$K=19$. \label{hze_fig3}}

\figcaption[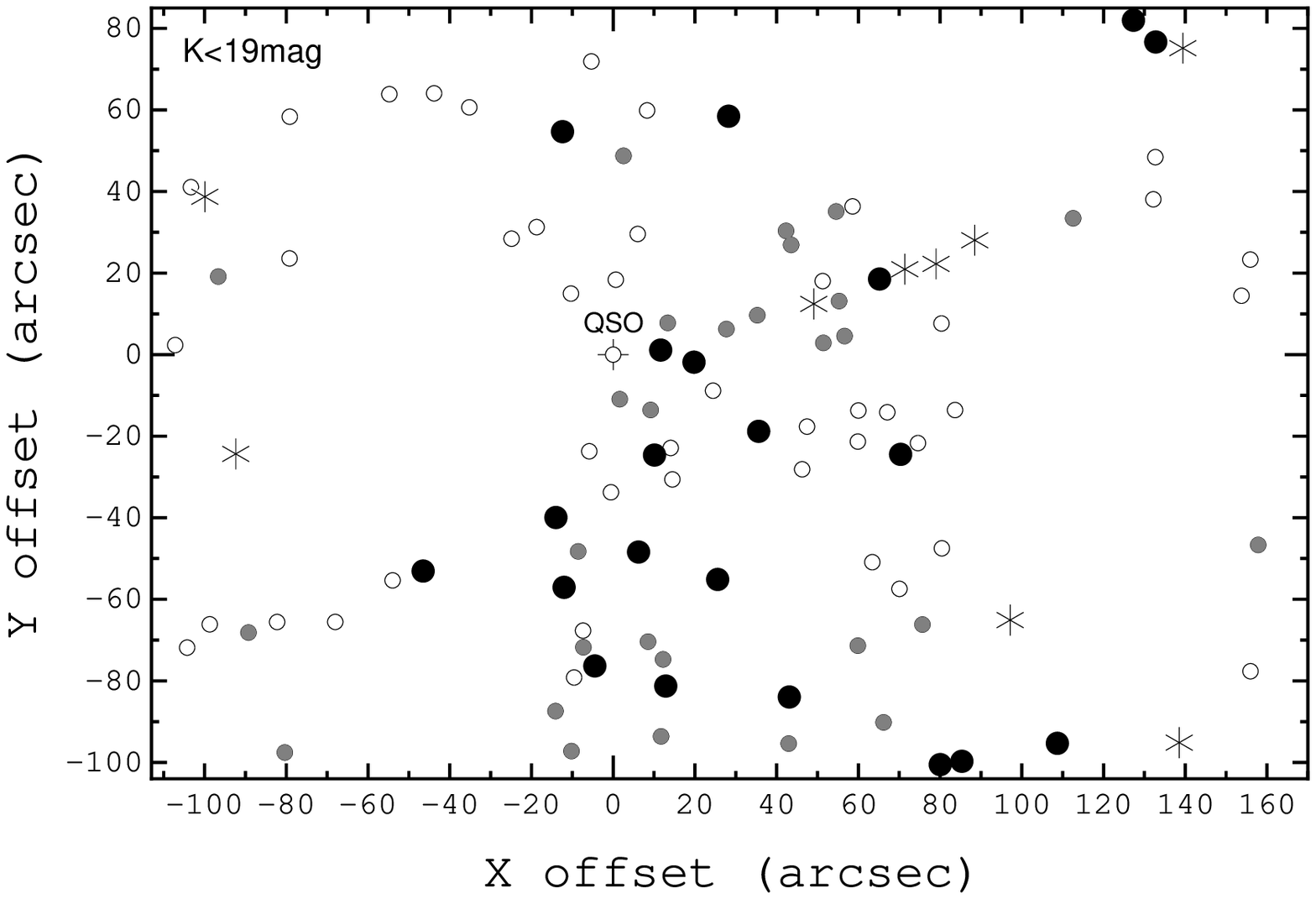]{The distribution of $K<19$ objects in the field.
Coordinates are offsets from the quasar in arcsecs. North is up and East
is left. Large filled circles, small filled circles and small open
circles represent the objects with $R-K>5$, $4<R-K<5$, and $R-K<4$,
respectively. Asterisks are objects that are consistent with stars
in both FWHM and colors. \label{hze_fig4}}

\figcaption[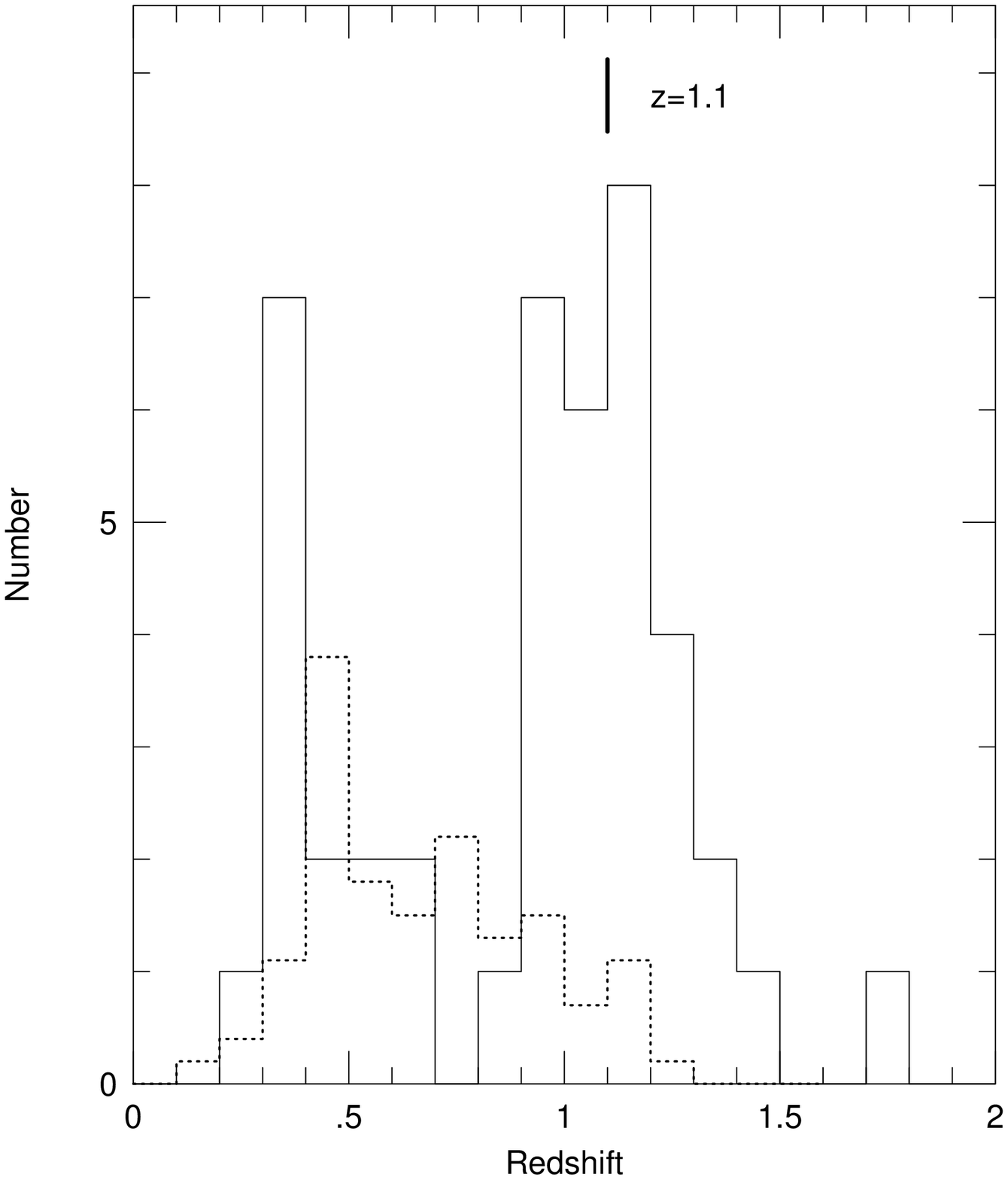]{Photometric redshift distribution
using Kodama, Bell \& Bower (1998) method.
We show all the objects in the Cluster region
($3^{\prime}\times1.5^{\prime}$ area) with $K<19\,$mag (solid line). As
a reference, the observed redshift distribution of $K<19$ field galaxies
from Cowie et al. (1996) is shown after scaling by area
(dashed line). \label{hze_fig5}}

\figcaption[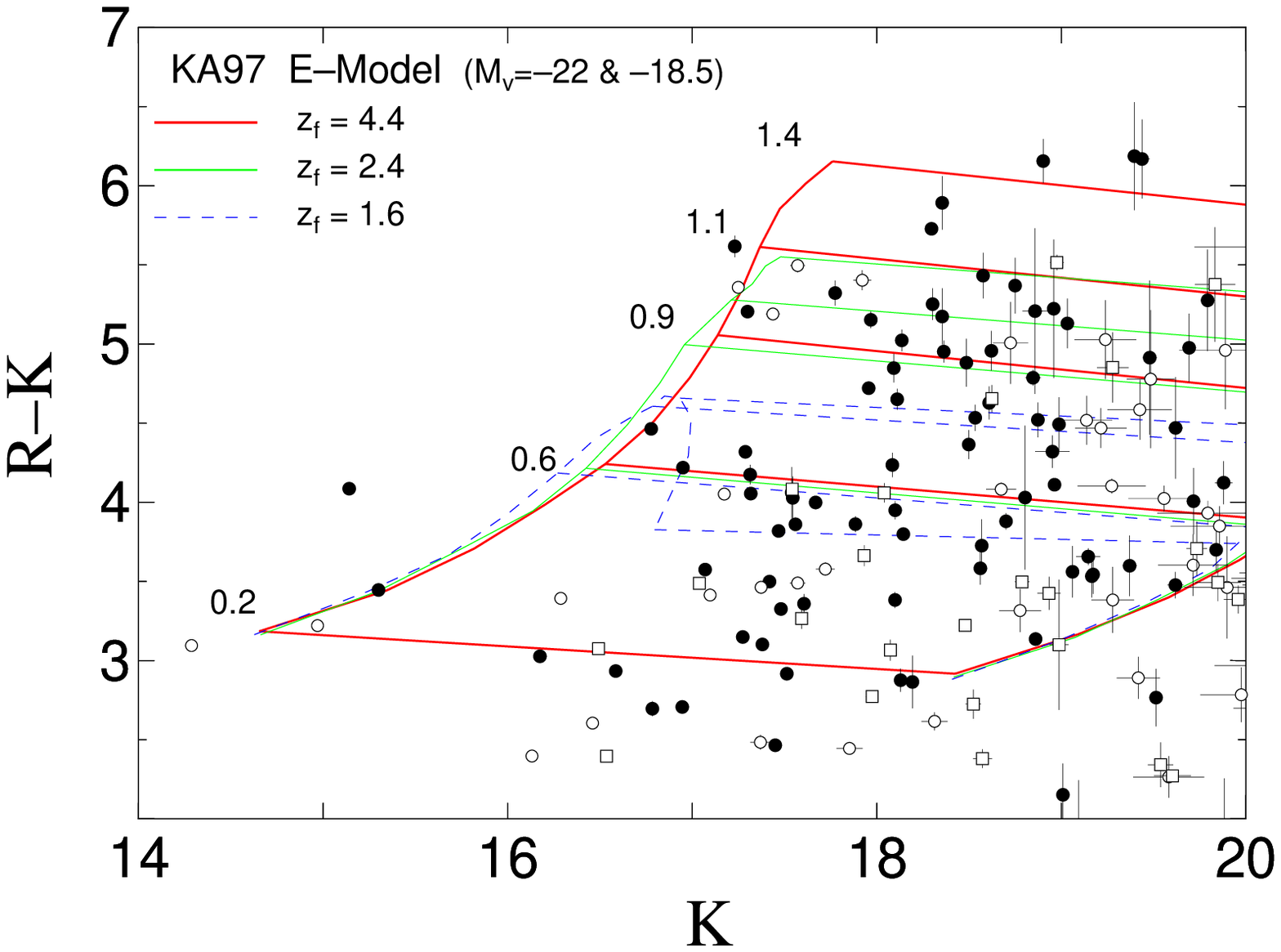]{$R-K$ versus  $K$ color-magnitude diagram for the
$K$-selected galaxies. Filled circles, open circles, and open boxes
represent the objects in the Cluster region, the Western region, and the
Eastern region, respectively (see Figure~\ref{hze_fig1}).
The model tracks shown as a thick solid line, thin solid line, and dashed
line are the ``Coma C-M model'' with
formation epochs ($z_{\rm f}$) of 4.4, 2.4 and 1.6 respectively
(see text, KA97). Evolutionary models with different masses ($M_{V}=-22$ and
$-18.5\,$mag at $z=0$) are shown.\label{hze_fig6}}

\figcaption[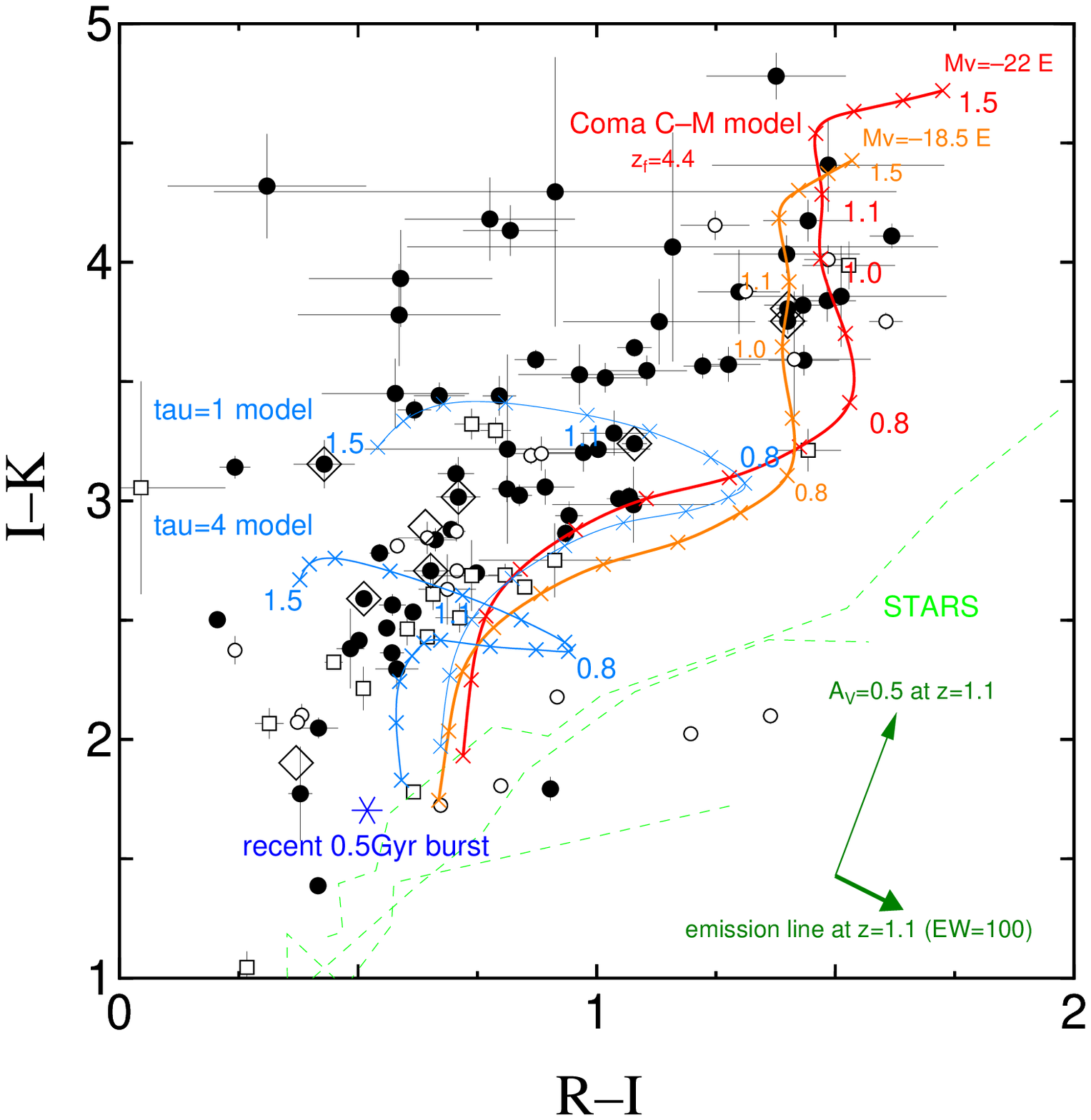]{$I-K$ versus $R-I$ two-color diagram
for all objects brighter than
$K=19$. The symbols are the same as in Figure \ref{hze_fig6}. The symbols
with large diamonds correspond to the emission-line galaxies found by
H93.  The model tracks shown as ``Coma C-M model'' are the same ones
used in Figure \ref{hze_fig6}. The exponentially-decaying SFR models with
time scales $\tau=1$ and 4~Gyr are also plotted. The asterisk shows the
color of a 0.5~Gyr constant star formation model observed at $z=1.1$.
The effect of internal extinction and that of an $EW=100$\AA\ emission
line in the $I$-band are shown as arrows in the bottom-right corner.
See text for details. \label{hze_fig7}}

\figcaption[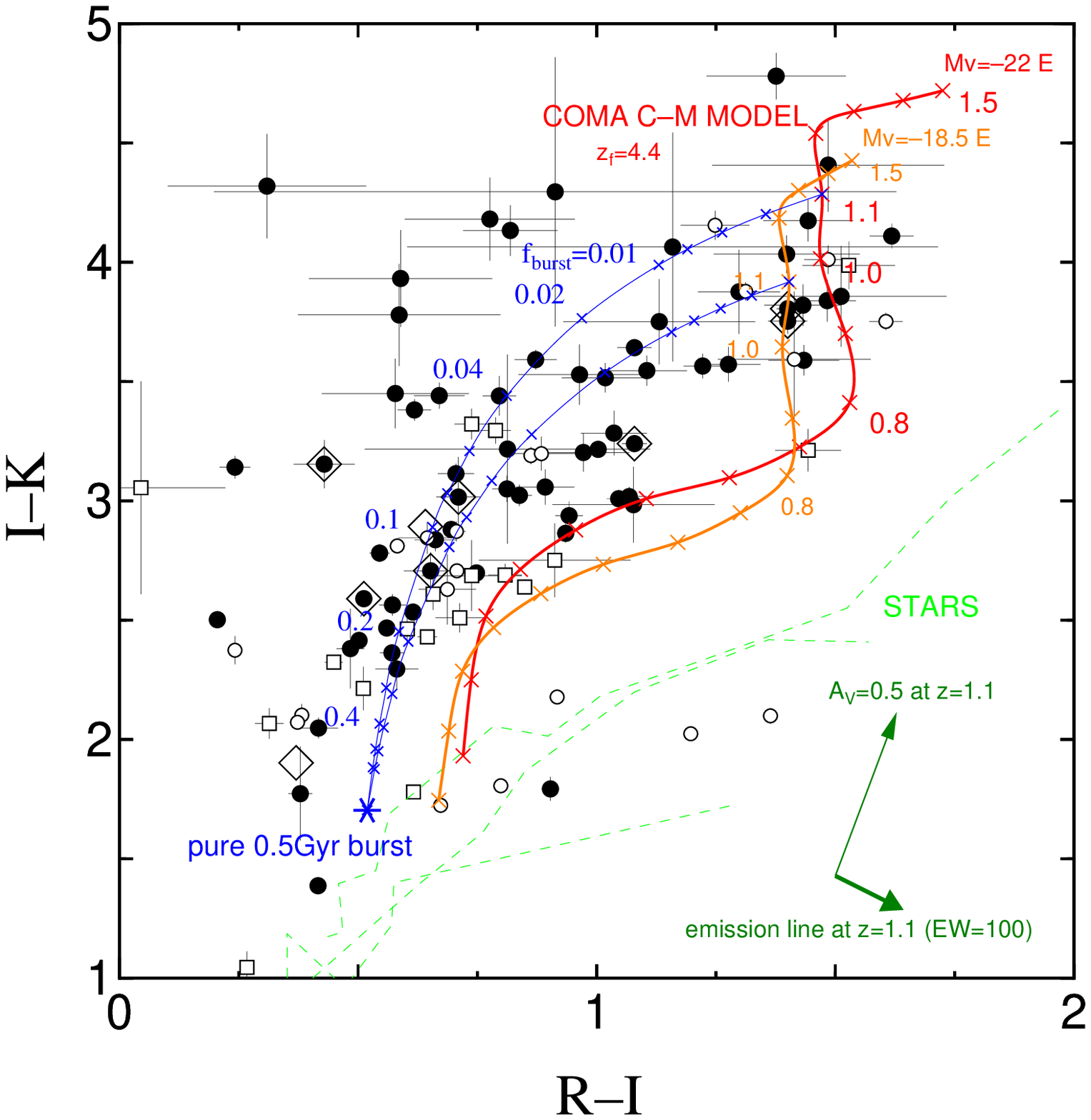]{The same as Figure~\ref{hze_fig7},
but with models in which various amounts
of an star-forming component are added to the passively-evolving old
component (see text). The passive models are shown again as thick
lines.  The old+star-forming tracks (thin lines) are for galaxies of
different metallicity/luminosity: the upper line is for a bright
metal-rich model and the lower for a faint metal-poor one following
the ``Coma C-M model'' of KA97.  The $f_{\rm burst}$ values indicate 
the mass fraction in the star-forming component (equation 1). \label{hze_fig8}}

\figcaption[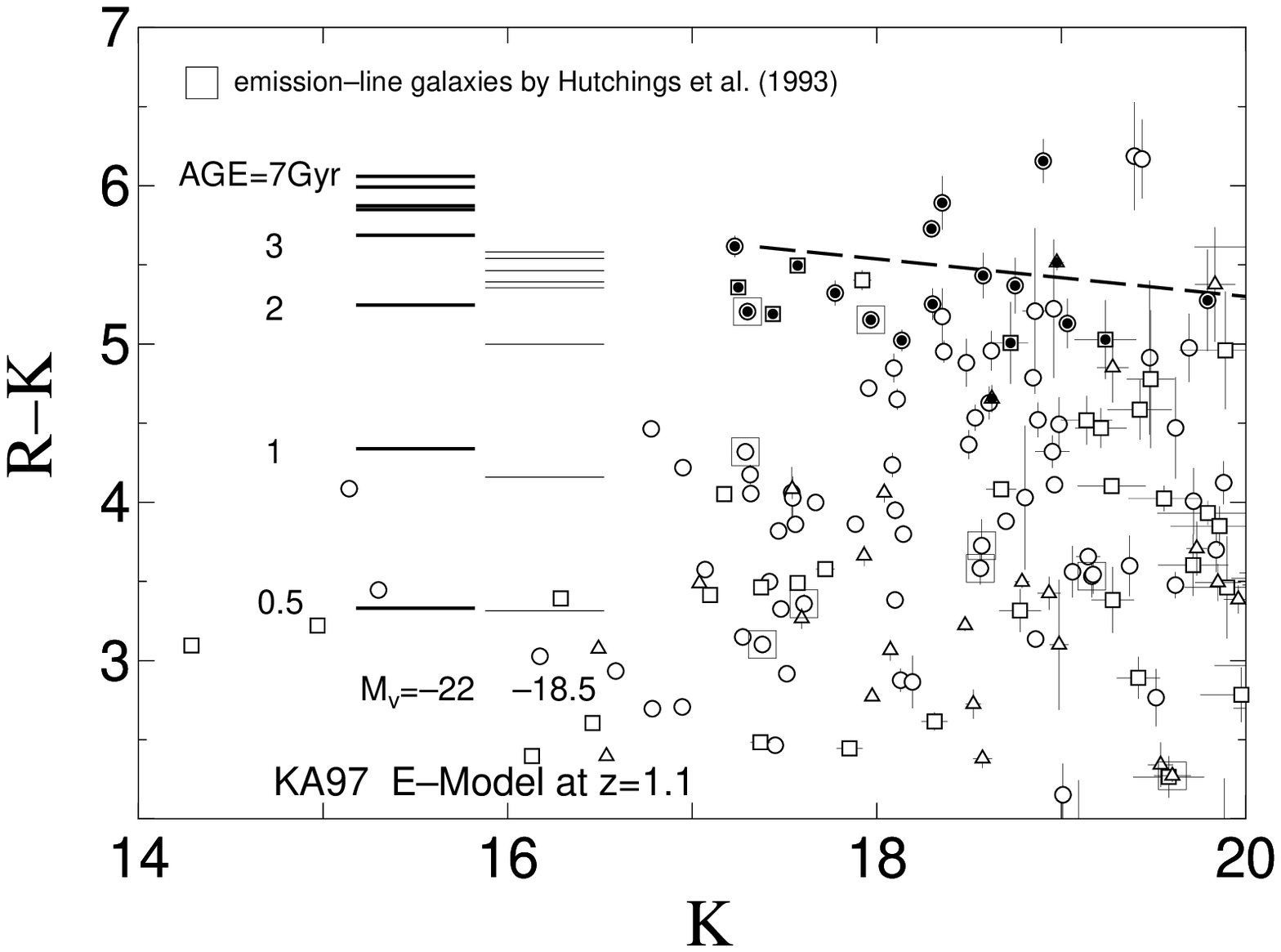]{The ``recovered'' C-M sequence for the
cluster galaxies (see text; 9a for $K$ vs $R-K$ and 9b for $K$
vs $I-K$). Circles, squares, and triangles represent galaxies
in the Cluster region, the Western region, and the Eastern region,
respectively. The symbols with small filled circles show the galaxies
having colors consistent with the passive evolution model and without
$UV$ excess. The dashed line shows the predicted C-M relation at $z=1.1$
($z_{f}=4.4$). The solid lines are the $z=1.1$ colors of the ``Coma C-M
model'' for ages between 0.5~Gyr and 7~Gyr (Note that the age of the
universe at $z=1.1$ is 4.3~Gyr in our cosmology).  The results for two
different mass/metallicity models (M$_{V}=-22$ and $-18.5$) are
shown.\label{hze_fig9}}

\figcaption[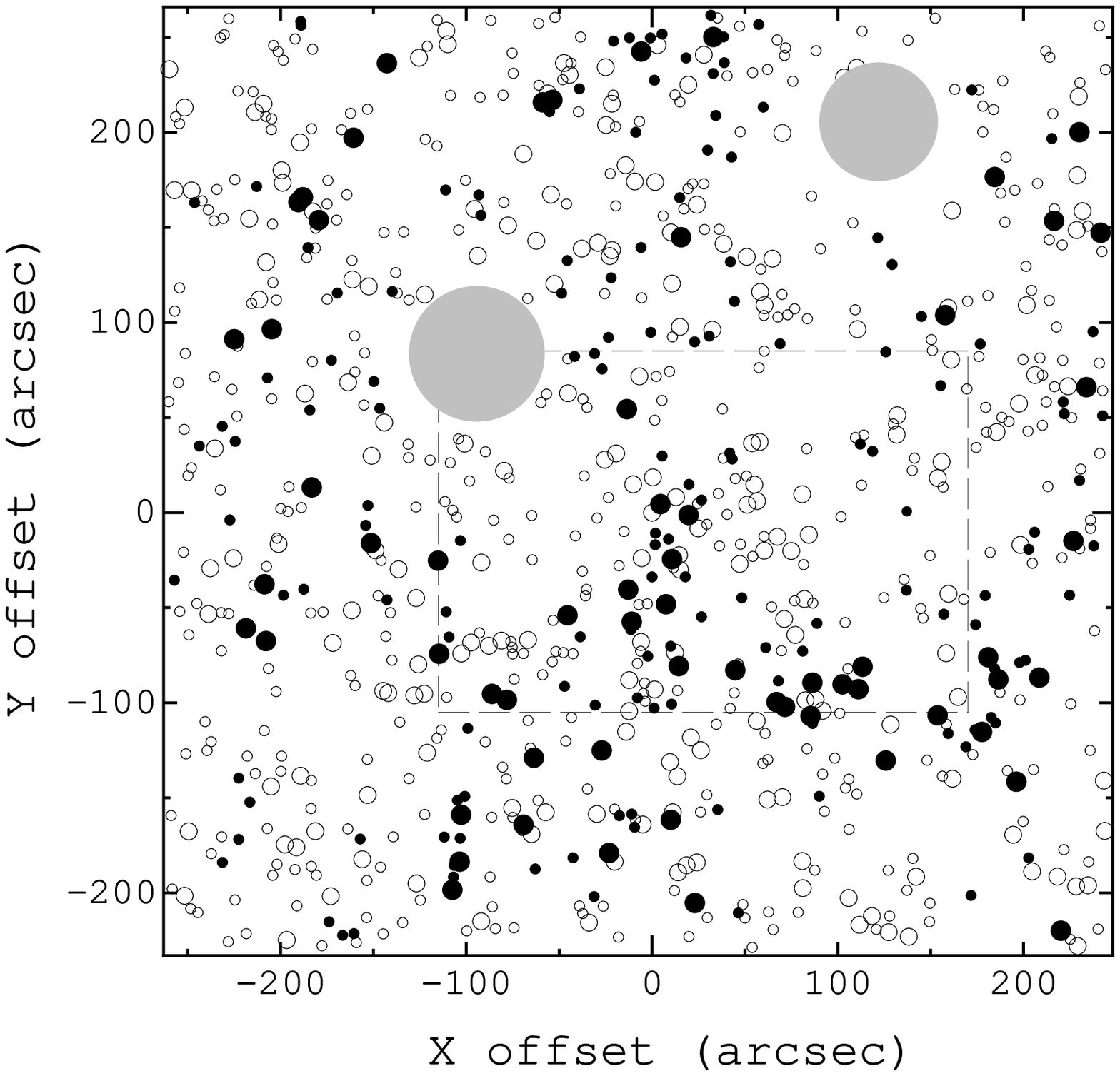]{The distribution of the $I<23.5$
galaxies on the CCD frame. A linear
scale of 100 arcsec corresponds to about 0.85~Mpc at the quasar redshift in our
adopted cosmology.  Filled circles are objects with $21<I<23.5$ and
$0.8<R-I<1.3$ (small filled circles) and $R-I>1.3$ (large filled circles).
Faint bluer objects ($21<I<23.5$ and $R-I<0.8$) are shown as small open
circles. The large open circles are bright ($I<21$) galaxies.  The
object coordinates are relative offsets from the quasar in
arcsec. The dashed rectangle shows the area of $K$-band frames. Two
large gray patches are the regions affected by bright stars and thus
excluded. Stars brighter than $I=21$ are removed, but fainter ones are
not removed because the star-galaxy separation is not robust for
fainter objects. \label{hze_fig10}}

\clearpage
\begin{figure}
\plotone{hze_fig1.eps}\\
Figure 1.

\end{figure}
\clearpage

\begin{figure}
Figure 2: COLOR IMAGE AVAILABLE AT {\tt http://xxx.lanl.gov/archive/astro-ph}
\end{figure}
\clearpage

\clearpage
\begin{figure}
\plotone{hze_fig3a.eps}\\
Figure 3a.
\end{figure}
\clearpage

\begin{figure}
\plotone{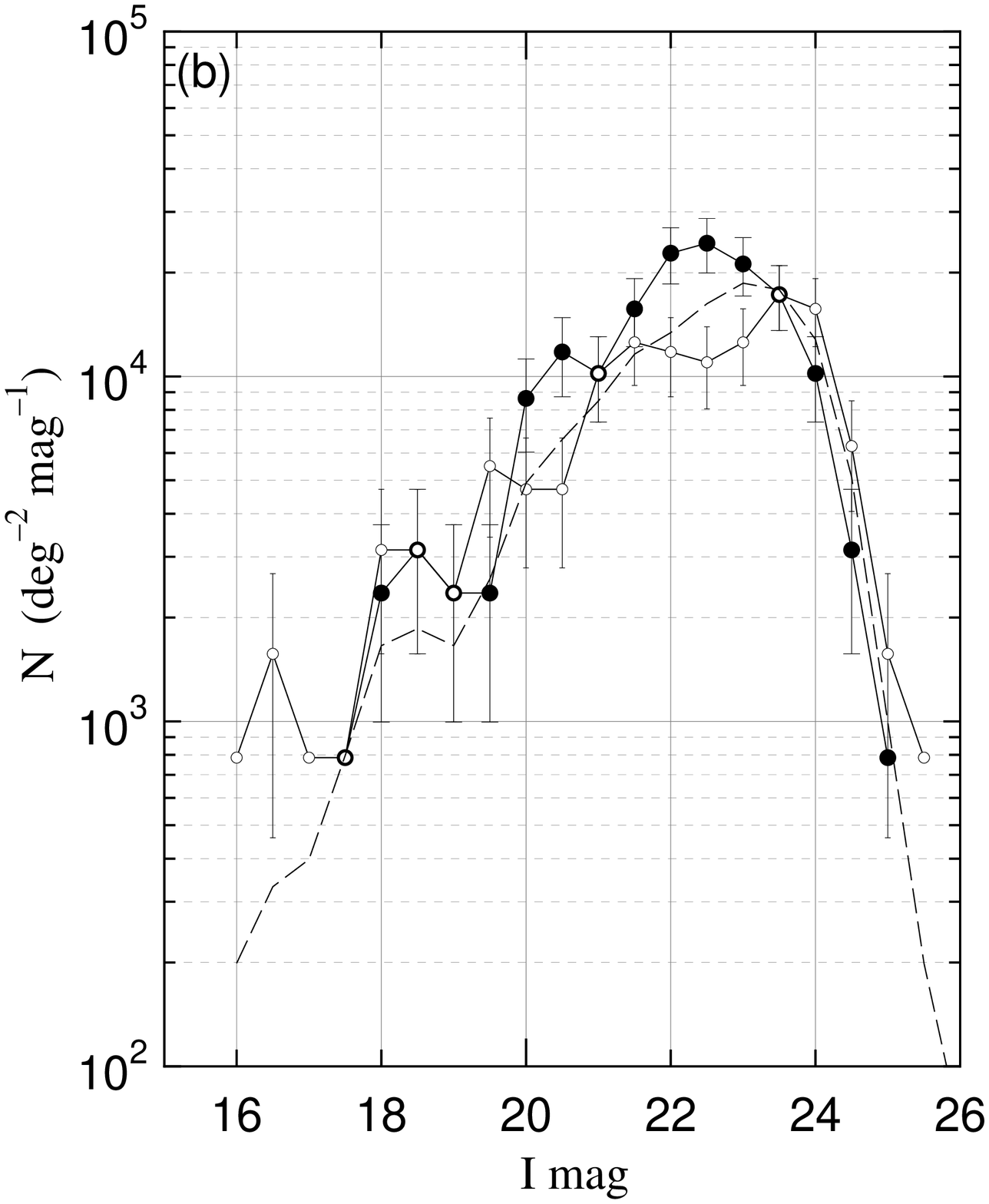}\\
Figure 3b.
\end{figure}
\clearpage

\begin{figure}
\plotone{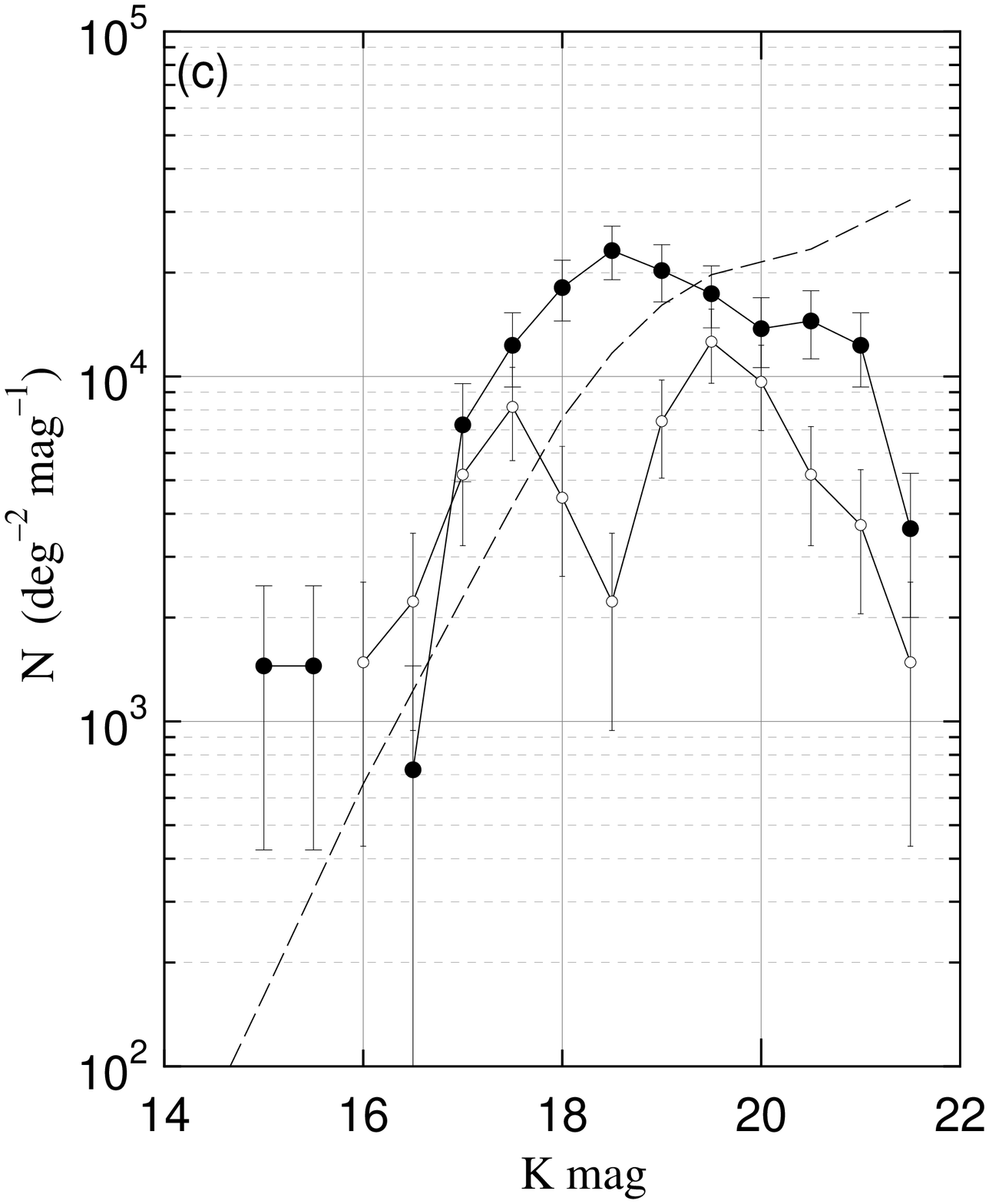}\\
Figure 3c.
\end{figure}
\clearpage

\begin{figure}
\plotone{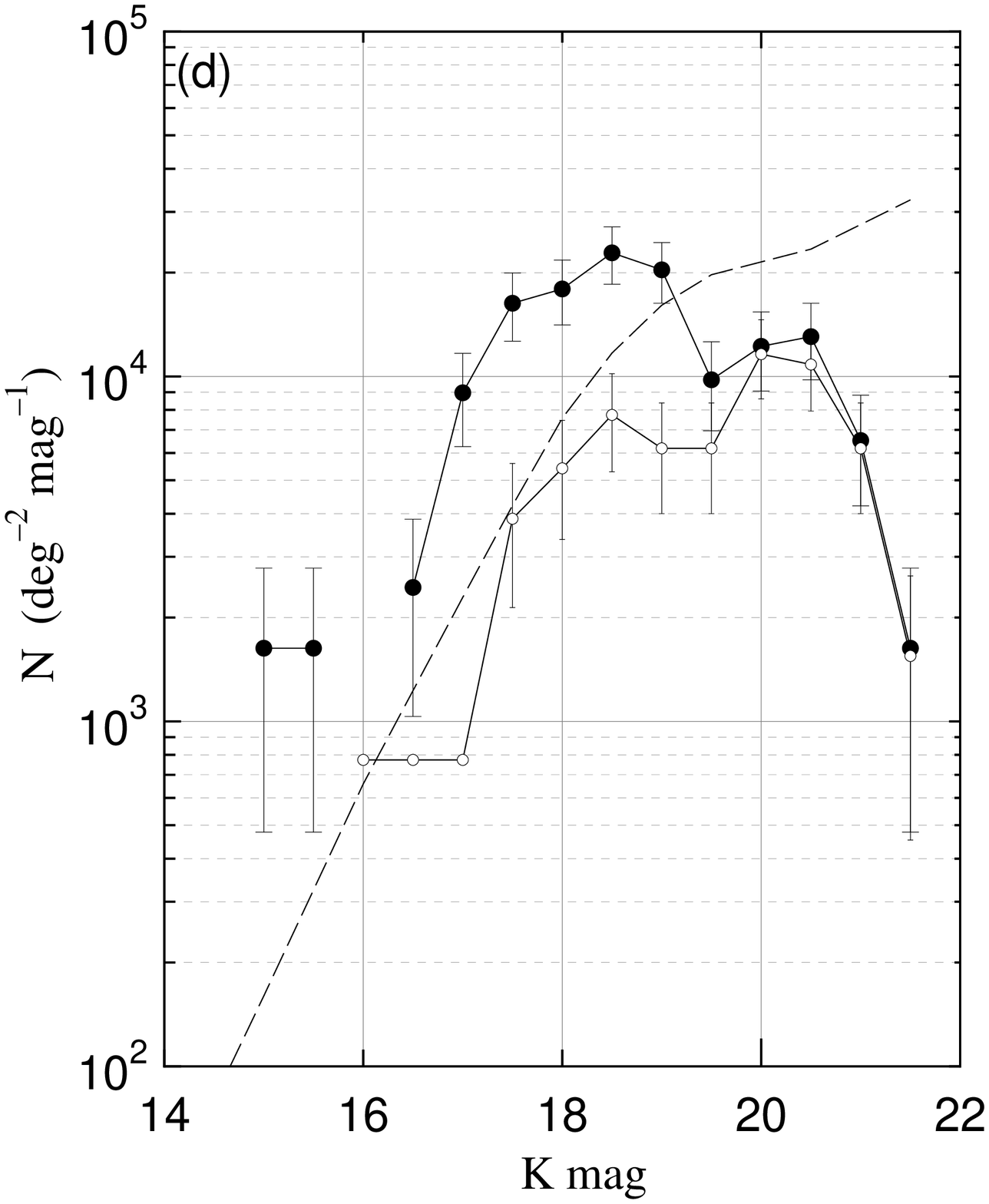}\\
Figure 3d.
\end{figure}
\clearpage

\begin{figure}
\plotone{hze_fig4.eps}\\
Figure 4.
\end{figure}
\clearpage

\begin{figure}
\plotone{hze_fig5.eps}\\
Figure 5.
\end{figure}
\clearpage

\begin{figure}
\plotone{hze_fig6.eps}\\
Figure 6.
\end{figure}
\clearpage

\begin{figure}
\plotone{hze_fig7.eps}\\
Figure 7.
\end{figure}
\clearpage

\begin{figure}
\plotone{hze_fig8.eps}\\
Figure 8.
\end{figure}
\clearpage

\begin{figure}
\plotone{hze_fig9a.eps}\\
Figure 9a.
\end{figure}
\clearpage

\begin{figure}
\plotone{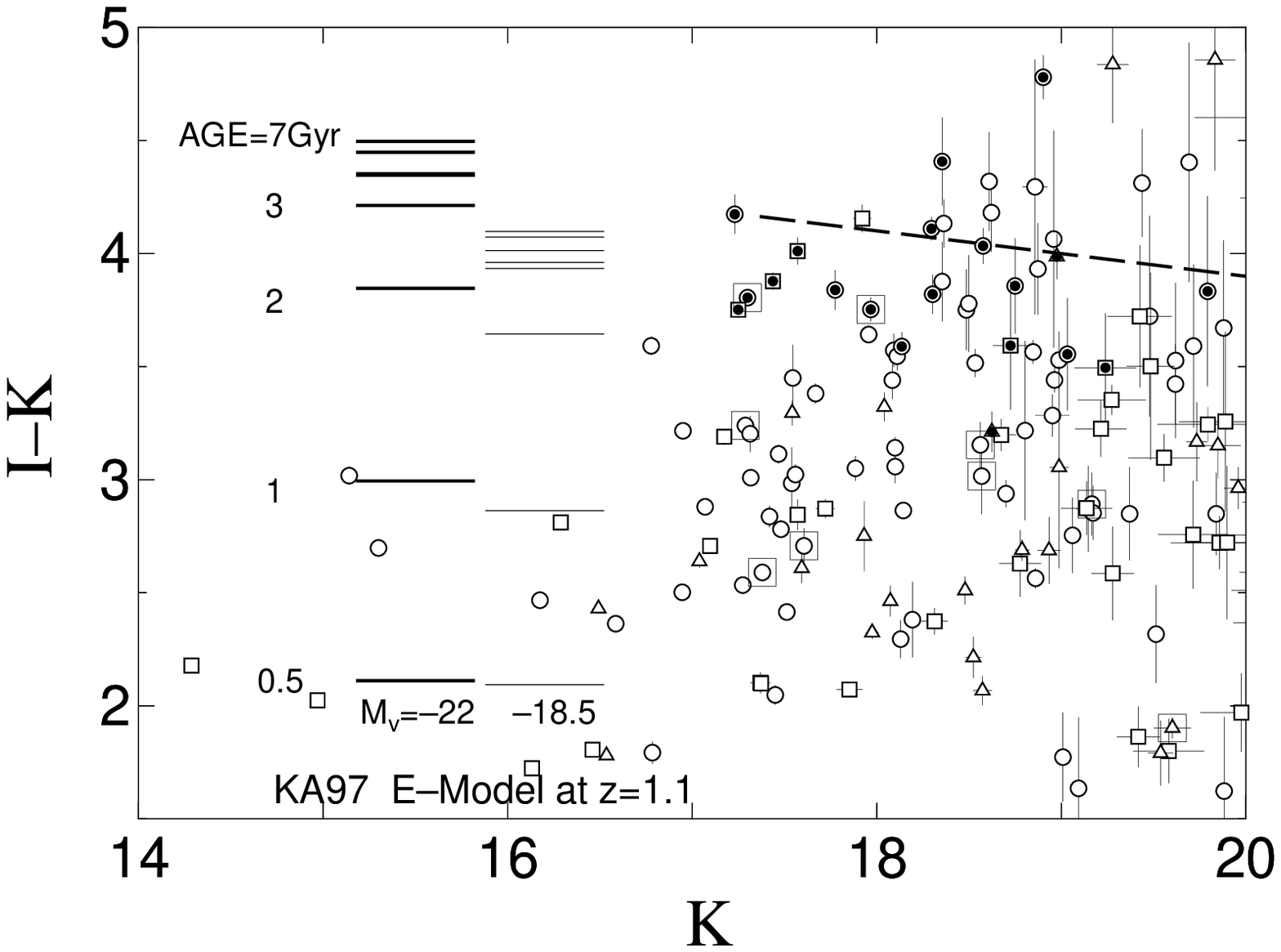}\\
Figure 9b.
\end{figure}
\clearpage

\begin{figure}
\plotone{hze_fig10.eps}\\
Figure 10.
\end{figure}
\clearpage
 
\newpage
\begin{deluxetable}{crrrrlrlrl}
\footnotesize
\tablecaption{$K$-selected object catalog. \label{tbl-1}}
\tablewidth{0pt}
\tablehead{
\colhead{ID\tablenotemark{a}} & \colhead{$X_{QSO}''$\tablenotemark{b}}
&\colhead{$Y_{QSO}''$\tablenotemark{b}}
&\colhead{$K_{\rm MB}$\tablenotemark{c}}
& \colhead{$R-I$\tablenotemark{d}} & \colhead{error}
& \colhead{$R-K$\tablenotemark{d}} & \colhead{error}
& \colhead{$I-K$\tablenotemark{d}} & \colhead{error}
}
\startdata
  1001 &  300.2 &   24.2 &  15.14 &   1.07 &   0.04 &   4.09 &    0.04 &    3.02 &   0.04 \nl
  1002 &  317.4 & -113.1 &  15.30 &   0.75 &   0.02 &   3.45 &    0.02 &    2.70 &   0.02 \nl
  1003 &  355.6 &  -74.9 &  16.17 &   0.56 &   0.01 &   3.03 &    0.02 &    2.47 &   0.02 \nl
  1004 & -131.9 &  150.7 &  16.59 &   0.57 &   0.03 &   2.93 &    0.02 &    2.36 &   0.03 \nl
  1005 &  401.2 & -350.9 &  16.78 &   0.87 &   0.04 &   4.46 &    0.03 &    3.59 &   0.04 \nl
  1006 &  260.3 &   66.0 &  16.78 &   0.90 &   0.02 &   2.70 &    0.05 &    1.79 &   0.05 \nl
  1007 &    0.0 &    0.0 &  16.95 &   0.21 &   0.01 &   2.71 &    0.03 &    2.50 &   0.03 \nl
  1008 &   70.7 &   41.3 &  16.95 &   1.00 &   0.01 &   4.22 &    0.03 &    3.22 &   0.03 \nl
  1009 &  245.2 & -149.1 &  17.07 &   0.70 &   0.01 &   3.58 &    0.02 &    2.88 &   0.02 \nl
  1010 &  378.4 &  111.1 &  17.20 &   0.42 &   0.01 &   1.80 &    0.02 &    1.39 &   0.02 \nl
  1011 &  104.8 &   -9.8 &  17.23 &   1.44 &   0.09 &   5.62 &    0.07 &    4.17 &   0.09 \nl
  1012 &  395.4 & -115.0 &  17.27 &   0.62 &   0.02 &   3.15 &    0.02 &    2.53 &   0.02 \nl
  1013 &  -74.9 & -463.4 &  17.29 &   1.08 &   0.03 &   4.32 &    0.03 &    3.24 &   0.03 \nl
  1014 &  -74.4 & -211.7 &  17.30 &   1.40 &   0.02 &   5.21 &    0.02 &    3.81 &   0.02 \nl
  1015 &  272.7 &   15.2 &  17.31 &   0.97 &   0.10 &   4.18 &    0.06 &    3.20 &   0.08 \nl
  1016 &  289.3 &  186.2 &  17.32 &   1.05 &   0.03 &   4.06 &    0.03 &    3.01 &   0.02 \nl
  1017 &  -28.3 &  381.3 &  17.38 &   0.51 &   0.02 &   3.10 &    0.03 &    2.59 &   0.03 \nl
  1018 &  -54.8 &   79.4 &  17.42 &   0.66 &   0.04 &   3.50 &    0.02 &    2.84 &   0.05 \nl
  1019 &    3.4 &   97.5 &  17.45 &   0.42 &   0.04 &   2.47 &    0.04 &    2.05 &   0.04 \nl
  1020 &  -99.4 &  165.8 &  17.47 &   0.71 &   0.04 &   3.82 &    0.02 &    3.11 &   0.04 \nl
  1021 &  -39.2 & -358.8 &  17.48 &   0.55 &   0.02 &   3.33 &    0.04 &    2.78 &   0.04 \nl
  1022 &  -31.0 & -125.6 &  17.51 &   0.50 &   0.01 &   2.92 &    0.04 &    2.42 &   0.04 \nl
  1023 &  293.0 &   69.7 &  17.54 &   1.08 &   0.17 &   4.06 &    0.16 &    2.98 &   0.16 \nl
  1024 &   64.7 & -396.2 &  17.54 &   0.58 &   0.15 &   4.03 &    0.14 &    3.45 &   0.15 \nl
  1025 &  310.6 &  192.7 &  17.56 &   0.84 &   0.03 &   3.86 &    0.04 &    3.02 &   0.05 \nl
  1026 &   76.8 & -162.2 &  17.61 &   0.65 &   0.06 &   3.36 &    0.06 &    2.71 &   0.08 \nl
  1027 &  -50.8 & -419.7 &  17.67 &   0.62 &   0.04 &   4.00 &    0.03 &    3.38 &   0.05 \nl
  1028 &   32.9 & -256.6 &  17.77 &   1.48 &   0.06 &   5.32 &    0.08 &    3.84 &   0.09 \nl
  1029 &  129.5 &  -46.9 &  17.88 &   0.81 &   0.03 &   3.86 &    0.05 &    3.05 &   0.06 \nl
  1030 &  350.8 & -478.1 &  17.96 &   1.08 &   0.04 &   4.72 &    0.04 &    3.64 &   0.03 \nl
  1031 &   53.6 & -130.5 &  17.97 &   1.40 &   0.04 &   5.15 &    0.05 &    3.75 &   0.05 \nl
  1032 &   48.6 &  -71.9 &  18.08 &   0.80 &   0.04 &   4.24 &    0.08 &    3.44 &   0.08 \nl
  1033 &  224.3 &  161.0 &  18.09 &   1.28 &   0.07 &   4.85 &    0.09 &    3.57 &   0.07 \nl
  1034 &   44.0 &  317.5 &  18.10 &   0.24 &   0.03 &   3.38 &    0.05 &    3.14 &   0.05 \nl
  1035 &   31.8 &  156.8 &  18.10 &   0.89 &   0.06 &   3.95 &    0.06 &    3.06 &   0.07 \nl
  1036 &  317.4 & -378.4 &  18.11 &   1.11 &   0.08 &   4.65 &    0.07 &    3.55 &   0.06 \nl
  1037 &  271.7 &   95.7 &  18.13 &   0.58 &   0.05 &   2.88 &    0.07 &    2.30 &   0.08 \nl
  1038 &  228.4 & -445.1 &  18.13 &   1.43 &   0.07 &   5.02 &    0.07 &    3.59 &   0.06 \nl
  1039 &   74.6 & -121.3 &  18.14 &   0.94 &   0.02 &   3.80 &    0.04 &    2.86 &   0.04 \nl
  1040 &  371.4 & -304.5 &  18.19 &   0.48 &   0.03 &   2.87 &    0.17 &    2.38 &   0.17 \nl
  1041 &  -65.8 &  289.9 &  18.30 &   1.62 &   0.05 &   5.73 &    0.04 &    4.11 &   0.05 \nl
  1042 &   68.2 & -430.9 &  18.30 &   1.43 &   0.08 &   5.25 &    0.10 &    3.82 &   0.09 \nl
  1043 &  135.6 & -292.3 &  18.35 &   1.30 &   0.09 &   5.17 &    0.18 &    3.88 &   0.18 \nl
  1044 &  -63.8 & -302.6 &  18.35 &   1.49 &   0.24 &   5.89 &    0.17 &    4.41 &   0.20 \nl
  1045 &   62.1 & -496.3 &  18.36 &   0.82 &   0.10 &   4.95 &    0.07 &    4.13 &   0.11 \nl
  1046 &  -54.2 & -515.5 &  18.49 &   1.13 &   0.20 &   4.88 &    0.15 &    3.75 &   0.18 \nl
  1047 &  -45.5 & -255.9 &  18.50 &   0.59 &   0.21 &   4.37 &    0.09 &    3.78 &   0.21 \nl
  1048 &   45.1 & -373.1 &  18.53 &   1.02 &   0.09 &   4.53 &    0.08 &    3.52 &   0.06 \nl
  1049 &  251.5 &  -93.6 &  18.56 &   0.43 &   0.06 &   3.58 &    0.10 &    3.15 &   0.10 \nl
  1050 &  318.1 &  -72.6 &  18.57 &   0.71 &   0.05 &   3.73 &    0.17 &    3.02 &   0.17 \nl
  1051 &   61.6 &    6.0 &  18.58 &   1.40 &   0.15 &   5.43 &    0.14 &    4.03 &   0.08 \nl
  1052 &  186.9 &   51.2 &  18.61 &   0.31 &   0.21 &   4.63 &    0.10 &    4.32 &   0.22 \nl
  1053 &  188.7 &  -99.6 &  18.62 &   0.78 &   0.18 &   4.96 &    0.13 &    4.18 &   0.17 \nl
  1054 &   -2.9 & -178.9 &  18.70 &   0.94 &   0.03 &   3.88 &    0.05 &    2.94 &   0.06 \nl
  1055 &  345.6 &   98.3 &  18.75 &   1.51 &   0.22 &   5.37 &    0.18 &    3.86 &   0.21 \nl
  1056 &  -38.8 & -380.5 &  18.80 &   0.81 &   0.30 &   4.03 &    0.46 &    3.22 &   0.40 \nl
  1057 &  230.9 &  142.7 &  18.85 &   1.22 &   0.04 &   4.79 &    0.05 &    3.57 &   0.05 \nl
  1058 &  372.8 & -129.6 &  18.86 &   0.91 &   0.72 &   5.21 &    0.52 &    4.30 &   0.56 \nl
  1059 &  336.2 & -269.9 &  18.86 &   0.57 &   0.03 &   3.14 &    0.04 &    2.56 &   0.05 \nl
  1060 &   13.4 &  258.5 &  18.87 &   0.59 &   0.19 &   4.52 &    0.11 &    3.93 &   0.20 \nl
  1061 &  149.7 &  310.0 &  18.90 &   1.38 &   0.15 &   6.16 &    0.14 &    4.78 &   0.10 \nl
  1062 &    8.5 &  -57.8 &  18.95 &   1.04 &   0.07 &   4.32 &    0.10 &    3.28 &   0.09 \nl
  1063 &  -23.8 & -404.7 &  18.96 &   1.16 &   0.56 &   5.22 &    0.44 &    4.06 &   0.48 \nl
  1064 &  227.6 & -505.7 &  18.96 &   0.67 &   0.05 &   4.11 &    0.03 &    3.44 &   0.06 \nl
  1065 &  146.9 &   33.3 &  18.99 &   0.96 &   0.13 &   4.49 &    0.17 &    3.53 &   0.13 \nl
  2001 &  734.5 & -504.4 &  14.29 &   0.92 &   0.00 &   3.10 &    0.00 &    2.18 &   0.00 \nl
  2002 &  418.9 &  118.0 &  14.97 &   1.20 &   0.00 &   3.22 &    0.00 &    2.02 &   0.00 \nl
  2003 &  515.1 & -345.1 &  16.13 &   0.67 &   0.00 &   2.40 &    0.00 &    1.72 &   0.00 \nl
  2004 &  826.7 &  123.6 &  16.29 &   0.58 &   0.01 &   3.39 &    0.01 &    2.81 &   0.02 \nl
  2005 &  469.1 &  148.7 &  16.46 &   0.80 &   0.01 &   2.61 &    0.02 &    1.81 &   0.02 \nl
  2006 &  700.8 &  202.0 &  17.10 &   0.71 &   0.02 &   3.41 &    0.02 &    2.71 &   0.03 \nl
  2007 &  836.9 & -247.3 &  17.17 &   0.86 &   0.01 &   4.05 &    0.03 &    3.19 &   0.04 \nl
  2008 &  424.4 & -533.0 &  17.25 &   1.61 &   0.04 &   5.36 &    0.03 &    3.75 &   0.04 \nl
  2009 &  443.5 &  -72.0 &  17.37 &   0.38 &   0.01 &   2.48 &    0.05 &    2.10 &   0.05 \nl
  2010 &  739.1 &  398.5 &  17.37 &   1.36 &   0.01 &   3.46 &    0.03 &    2.10 &   0.03 \nl
  2011 &  703.9 &  406.5 &  17.44 &   1.31 &   0.02 &   5.19 &    0.04 &    3.88 &   0.04 \nl
  2012 &  826.9 & -411.6 &  17.57 &   0.65 &   0.06 &   3.49 &    0.04 &    2.85 &   0.07 \nl
  2013 &  576.3 & -505.3 &  17.57 &   1.49 &   0.05 &   5.50 &    0.04 &    4.01 &   0.06 \nl
  2014 &  425.9 &   40.5 &  17.72 &   0.71 &   0.02 &   3.58 &    0.04 &    2.87 &   0.04 \nl
  2015 &  426.4 & -251.8 &  17.85 &   0.37 &   0.01 &   2.45 &    0.01 &    2.07 &   0.02 \nl
  2016 &  675.0 &  434.8 &  17.92 &   1.25 &   0.07 &   5.40 &    0.06 &    4.16 &   0.06 \nl
  2017 &  703.4 &  256.8 &  18.31 &   0.24 &   0.02 &   2.62 &    0.06 &    2.37 &   0.06 \nl
  2018 &  596.8 &  177.4 &  18.67 &   0.88 &   0.07 &   4.08 &    0.04 &    3.20 &   0.07 \nl
  2019 &  452.4 & -528.7 &  18.72 &   1.41 &   0.16 &   5.01 &    0.26 &    3.59 &   0.28 \nl
  2020 &  815.1 &   76.6 &  18.78 &   0.69 &   0.06 &   3.32 &    0.14 &    2.63 &   0.15 \nl
  3001 & -552.8 & -380.8 &  16.49 &   0.65 &   0.02 &   3.08 &    0.02 &    2.43 &   0.02 \nl
  3002 & -489.7 & -129.1 &  16.53 &   0.62 &   0.01 &   2.40 &    0.02 &    1.78 &   0.02 \nl
  3003 & -436.1 & -347.6 &  17.04 &   0.85 &   0.01 &   3.49 &    0.03 &    2.64 &   0.03 \nl
  3004 & -473.2 & -361.3 &  17.54 &   0.79 &   0.03 &   4.08 &    0.06 &    3.30 &   0.06 \nl
  3005 & -360.7 & -347.5 &  17.59 &   0.66 &   0.01 &   3.27 &    0.07 &    2.61 &   0.07 \nl
  3006 & -529.8 &  205.3 &  17.79 &   0.27 &   0.00 &   1.31 &    0.07 &    1.05 &   0.07 \nl
  3007 & -419.7 &  309.4 &  17.93 &   0.91 &   0.16 &   3.66 &    0.07 &    2.75 &   0.16 \nl
  3008 & -523.5 & -350.7 &  17.97 &   0.45 &   0.02 &   2.77 &    0.03 &    2.32 &   0.03 \nl
  3009 & -512.4 &  101.5 &  18.04 &   0.74 &   0.04 &   4.06 &    0.06 &    3.32 &   0.06 \nl
  3010 & -232.4 &  339.8 &  18.07 &   0.60 &   0.02 &   3.07 &    0.07 &    2.46 &   0.07 \nl
  3011 & -186.7 &  321.5 &  18.48 &   0.71 &   0.05 &   3.22 &    0.04 &    2.51 &   0.06 \nl
  3012 & -547.9 &  217.8 &  18.52 &   0.51 &   0.01 &   2.73 &    0.09 &    2.21 &   0.09 \nl
  3013 & -420.0 &  125.1 &  18.57 &   0.31 &   0.03 &   2.38 &    0.06 &    2.07 &   0.06 \nl
  3014 & -426.1 & -517.3 &  18.62 &   1.44 &   0.07 &   4.66 &    0.09 &    3.21 &   0.09 \nl
  3015 & -286.2 & -293.5 &  18.79 &   0.81 &   0.04 &   3.50 &    0.04 &    2.69 &   0.05 \nl
  3016 & -290.5 &  338.7 &  18.93 &   0.74 &   0.11 &   3.43 &    0.10 &    2.69 &   0.15 \nl
  3017 & -246.7 & -281.4 &  18.98 &   1.53 &   0.10 &   5.52 &    0.05 &    3.99 &   0.10 \nl
  3018 & -568.3 &   12.4 &  18.99 &   0.05 &   0.18 &   3.10 &    0.41 &    3.06 &   0.45 \nl
\enddata
\tablenotetext{a}{ID numbers 1000s, 2000s, and 3000s 
 represent galaxies in the Cluster region, the Western region, and the
 Eastern region, respectively.}
\tablenotetext{b}{Relative offsets for each object from the quasar
 in arcsec.}
\tablenotetext{c}{The MAG\_BEST output for the $K$-band data produced by 
 SExtractor (Bertin \& Arnouts 1996). These values are used as total magnitudes
 for the galaxies.}
\tablenotetext{d}{Colors are measured in 3.5-arcsec-diameter apertures 
for the seeing-matched frames. The tabulated errors are
empirical values based on repeated photometry for each object 
(see Sect. 2).}
\end{deluxetable}

\end{document}